\pdfoutput=1
\documentclass[aps,prx,floatfix,twocolumn,superscriptaddress]{revtex4-1}  %superscriptaddress
\usepackage{physics}
\usepackage{amsfonts}
\usepackage{mathrsfs}
\usepackage{amsmath}% needed for subequations
\usepackage{multirow}% needed for multirow table
\usepackage{color}
\usepackage{graphicx}
\usepackage{bm}% bold maths
\usepackage{amssymb}
\usepackage{xspace}
\usepackage{epstopdf}
\usepackage{dcolumn}% Align table columns on decimal point
\usepackage{longtable}
\usepackage{float}
\usepackage[colorlinks=true, letterpaper=true, pdfstartview=FitV, linkcolor=blue, citecolor=blue, urlcolor=blue]{hyperref}
 
\usepackage{scalerel}

\begin{document}
\title{Quasi-One-Dimensional Higher-Order Topological Insulators} % Bismuth Halides
\author{Chiho Yoon}
\affiliation{Department of Physics, University of Texas at Dallas, Richardson, Texas 75080, USA}
\affiliation{Department of Physics and Astronomy, Seoul National University, Seoul 08826, Korea}
\author{Cheng-Cheng Liu}
\affiliation{Department of Physics, University of Texas at Dallas, Richardson, Texas 75080, USA}
\affiliation{School of Physics, Beijing Institute of Technology, Beijing 100081, China}
\author{Hongki Min}\email{hmin@snu.ac.kr}
\affiliation{Department of Physics and Astronomy, Seoul National University, Seoul 08826, Korea}
\author{Zhang Fan}\email{zhang@utdallas.edu}
\affiliation{Department of Physics, University of Texas at Dallas, Richardson, Texas 75080, USA}

\begin{abstract}
%Quasi-1D materials Bi$_{4}$X$_{4}$~(X=Br,~I) are prototype weak topological insulators (TI) in the $\beta$ phase.  
%For the $\alpha$ phases, recent high-throughput database screening suggests that 
%Bi$_{4}$Br$_{4}$ is a rare higher-order TI (HOTI) whereas Bi$_{4}$I$_{4}$ has trivial symmetry indicators.   
%Here we show that in fact the two $\alpha$ phases are both pristine HOTIs yet with distinct hinge state patterns. 
%The location of inversion center dictates Bi$_{4}$Br$_{4}$ (Bi$_{4}$I$_{4}$) to feature  
%opposite (the same) dimerizations at two side cleavage surfaces. 
%A variety of experiments are proposed to examine our predictions.
%Given the superior hinges along atomic chains,  
%the structural transition at room temperature, and the extreme anisotropies in three axes,
%our results not only imply the possible existence of many topological materials beyond the scope of symmetry indicators 
%but also establish a new TI paradigm and a unique material platform for exploring the interplay of geometry, symmetry, topology, and interaction.
% 
Quasi-1D materials Bi$_{4}$X$_{4}$~(X=Br,\,I) are prototype weak topological insulators (TI) in the $\beta$ phase.  
For the $\alpha$ phases, recent high-throughput database screening suggests that 
Bi$_{4}$Br$_{4}$ is a rare higher-order TI (HOTI) whereas Bi$_{4}$I$_{4}$ has trivial symmetry indicators.   
Here we show that in fact the two $\alpha$ phases are both pristine HOTIs yet with distinct termination-dependent hinge state patterns  
by performing first-principles calculations, analyzing coupled-edge dimerizations,  
inspecting surface lattice structures, constructing tight-binding models, and establishing boundary topological invariants. 
We reveal that the location of inversion center dictates Bi$_{4}$Br$_{4}$ (Bi$_{4}$I$_{4}$) to feature  
opposite (the same) dimerizations of a surface or intrinsic (bulk or extrinsic) origin at two side cleavage surfaces. 
We propose a variety of experiments to examine our predictions. 
Given the superior hinges along atomic chains,  
the structural transition at room temperature, and the extreme anisotropies in three axes,
our results not only imply the possible existence of many topological materials beyond the scope of symmetry indicators 
but also establish a new TI paradigm and a unique material platform for exploring the interplay of geometry, symmetry, topology, and interaction.
\end{abstract}
\date{\today}
\maketitle

\section{Introduction}

Geometry, symmetry, topology, and interaction are fundamental themes in physics. 
Their interplay governs microscopic laws of individual particles and macroscopic phenomena 
of many-particle systems. As a paradigm in condensed matter physics, 
Kane-Mele topological insulators (TI) exist in 2D and 3D but not in 0D and 1D, 
dictated by the time-reversal ($\mathcal{T}$) 
and gauge symmetries in spin-orbit-coupled systems~\cite{Kane2005, Fu2007a, Moore2007, Roy2009}.
For a 3D (2D) TI, the nontrivial $\mathbb{Z}_2$ topological invariant of a gapped bulk state 
implies the presence of a symmetry-protected gapless 2D (1D) surface (edge) state.
Having taken the electronics community by storm, 
not only has this spirit topologically classified all insulators, (semi-)metals, and superconductors 
with various different symmetries and dimensions~\cite{Hasan2010, Qi2011, Chiu2016, Armitage2018}, 
but it has also substantially inspired the study of classical systems that address photonic, acoustic, mechanical, 
and even equatorial waves~\cite{Zhang2018, Lu2016, Lu2017, Huber2016, Delplace2017}.
  
Fascinatingly, higher-order TIs (HOTI) and topological superconductors have emerged 
recently~\cite{Zhang2013, Benalcazar2017, Langbehn2017, Song2017, Schindler2018, Ezawa2018, Schindler2018a, Khalaf2018, Wang2018, Yan2018}. 
They host protected gapless states at boundaries of more than one dimension lower. 
In fact, they can be best exemplified by simple models  
of $\mathbb{Z}_2$ TIs under suitable symmetry breaking.
In the original work by Zhang, Kane, and Mele (ZKM)~\cite{Zhang2013}, when the $\mathcal{T}$ symmetry is broken for a 3D TI, 
one chiral hinge mode propagates along any hinge that reverses the sign of Hall conductivity of two magnetic gapped surfaces. 
In more recent studies, when the gauge symmetry is broken for a 2D TI, two Majorana corner modes are bound to any corner 
that reverses the sign of induced pairing of two superconducting gapped edges~\cite{Wang2018, Yan2018}.
While the former may be realized in Sm-doped Bi$_2$Se$_3$ or MnBi$_{2n}$Te$_{3n+1}$ in a magnetic field,
its thin-film limit as the first quantum anomalous Hall effect has been achieved in Cr-doped (Bi,Sb)$_2$Te$_3$~\cite{Yue2019, Zhang2020, Yu2010, Chang2013}.
While the latter may be materialized in the 112-family of iron pnictides such as Ca$_{1-x}$La$_x$FeAs$_2$, 
its 3D counterpart with helical Majorana hinge modes has appeared to be confirmed in FeTe$_{0.55}$Se$_{0.45}$~\cite{Wu2020, Zhang2019a, Gray2019}.
These progresses urge a more ambitious question: 
does there exist any {\em pristine} HOTI, i.e., a material with a {\em global} bulk band gap 
and a {\em natural} gapless hinge state in the {\em absence} of any symmetry-breaking perturbation? 

Following the theory of topological quantum chemistry or symmetry indicators~\cite{Bradlyn2017, Kruthoff2017, Po2017, Song2018, Khalaf2018a}, 
high-throughput screening of nonmagnetic topological materials has been performed in the Inorganic Crystal Structure Database~\cite{Hellenbrandt2004}, 
and thousands of candidates have been identified~\cite{Zhang2019, Vergniory2019, Tang2019}.
Unfortunately, only a handful of them are HOTIs with helical hinge states~\footnote{Hongming Weng (private communication). 
According to the high-throughput screening calculations~\cite{Zhang2019}, the list of most plausible candidates 
for HOTIs with global band gaps and helical hinge states includes $\alpha$-Bi$_{4}$Br$_{4}$ (but not $\alpha$-Bi$_{4}$I$_{4}$), 
KHgSb (with hourglass surface states), Bernal graphite (with a band gap $\sim0.025$~meV), 
IV-VI semiconductors (with Dirac surface states) and BaTe in rocksalt structure,  
ThTaN$_{3}$ in perovskite structure, La$_{2}$Hf$_{2}$O$_{7}$ in pyrocholore structure, 
and antiperovskite oxides A$_{3}$BO (A = Ca, Sr, Ba, Yb and B = Ge, Sn, Pb)}.  
Even though the list is short, extra efforts including more accurate calculations and more physical understanding are necessary 
to narrow down it to those not only truly topological but also experimentally feasible~\cite{Zunger2019}.  
As a rare yet prime example, it has been predicted that $\alpha$-Bi$_{4}$Br$_{4}$ is a HOTI with $\mathbb{Z}_4=2$ 
whereas $\alpha$-Bi$_{4}$I$_{4}$ has completely trivial symmetry indicators~\cite{Zhang2019, Vergniory2019, Tang2019, Tang2019a, Hsu2019}. 
Here, by using various of different computational and analytical approaches, 
we explicitly demonstrate that both $\alpha$-Bi$_{4}$Br$_{4}$ and $\alpha$-Bi$_{4}$I$_{4}$ 
are HOTIs with helical hinge states, though with sharp distinction in their hinge state patterns.
Given that the symmetry indicators of $\alpha$-Bi$_{4}$I$_{4}$ are indeed trivial, 
significantly, our results imply that there are likely to be many topological materials 
beyond the scope of symmetry indicators and awaiting to be discovered. 

Bi$_{4}$X$_{4}$~(X=Br,\,I) are quasi-1D van der Waals materials,    
and each can be viewed as a periodic stack of atomic chains.
The $\beta$ phase has been predicted as a prototype weak TI (WTI)~\cite{Liu2016}, 
as confirmed by angle-resolved photoemission spectroscopy (ARPES)~\cite{Noguchi2019}.  
Strain can further tune the $\beta$ phase between WTI, strong TI (STI), and normal insulator~\cite{Liu2016}.
Superior to layered WTIs, the quasi-1D WTI is granted two natural cleavage surfaces 
in which the distinct surface hallmarks defining WTI can be inspected~\cite{Liu2016}.  
Here we show that our revealed hinge states of the two $\alpha$ phases propagate 
along the natural cleavage hinges, i.e., in the chain direction. 
This extraordinary property would greatly facilitate the experimental detections of the hinge states.
Interestingly, intrinsic superconductivity has already been reported for both Bi$_{4}$Br$_{4}$ 
and Bi$_{4}$I$_{4}$~\cite{Pisoni2017, Wang2018a, Qi2018, Li2019}. 
This enables possible topological superconductivity to arise from an intrinsic proximity effect.
More remarkably for Bi$_{4}$I$_{4}$, the structural transition turns out to be 
around 300~K~\footnotetext[101]{Bing Lv (private communication).}~\cite{VonSchnering1978, Dikarev2001, Weiz2017, Noguchi2019, Note101}. 
Thus, our identifying $\alpha$-Bi$_{4}$I$_{4}$ as a HOTI not only uncovers a 
thermal phase transition between the first- and second-order TIs
but also implies that the topological surface/hinge states of Bi$_{4}$I$_{4}$ can be switched at room temperature,  
a property that may be exploited for potential applications.

Besides demonstrating the existence and superiority of helical hinge states in the two $\alpha$ phases, 
we reveal that the location of inversion center, overlooked before, 
plays a critical role in determining the termination-dependent hinge state patterns. 
The inversion center of Bi$_{4}$Br$_{4}$ (Bi$_{4}$I$_{4}$) lies in its (001) monolayer (bilayer center).
As a consequence, Bi$_{4}$Br$_{4}$ (Bi$_{4}$I$_{4}$) is an intrinsic (extrinsic) HOTI, 
and its two side cleavage surfaces exhibit opposite (the same) dimerizations of a surface (bulk) origin. 
Compelling evidence for the two novel yet distinct band topologies  
is provided by our advanced first-principles calculations, coupled-edge layer constructions, 
surface lattice-structure inspections, effective tight-binding models, and nontrivial surface winding numbers. 
Never formulated before, our boundary topological invariant
not only validates but also distinguishes the two HOTIs. 
Fitting well with the {\it ab initio} results, 
our model elucidates the intimate relation between 2D TI, WTI, and the two HOTIs. 
To examine our predictions, we point out how a variety of experiments can be carried out.
Our results establish a new TI physics paradigm and a unique quasi-1D material platform 
for exploring the interplay of geometry, symmetry, topology, and interaction. 

\section{Structure and Termination Dependent Helical Hinge States}\label{hinge}

In this section, we present the most striking features of 
$\alpha$-Bi$_{4}$Br$_{4}$ and $\alpha$-Bi$_{4}$I$_{4}$ as HOTIs, 
i.e., the emergence of helical hinge states and their dependence on the (001) termination.  
In order to display the hinge states, 
we build the {\it ab initio} tight-binding models for both materials  
by using the maximally localized Wannier functions (MLWF) for the $p$ orbitals of Bi and Br/I. 
The MLWF are constructed from the bulk density functional theory (DFT) results 
that are obtained by the Heyd-Scuseria-Ernzerhof (HSE) hybrid functional. 
(The computational methods are detailed in Sec.~\ref{sec:band}.)

\begin{figure*}[t!]
\includegraphics[scale=1]{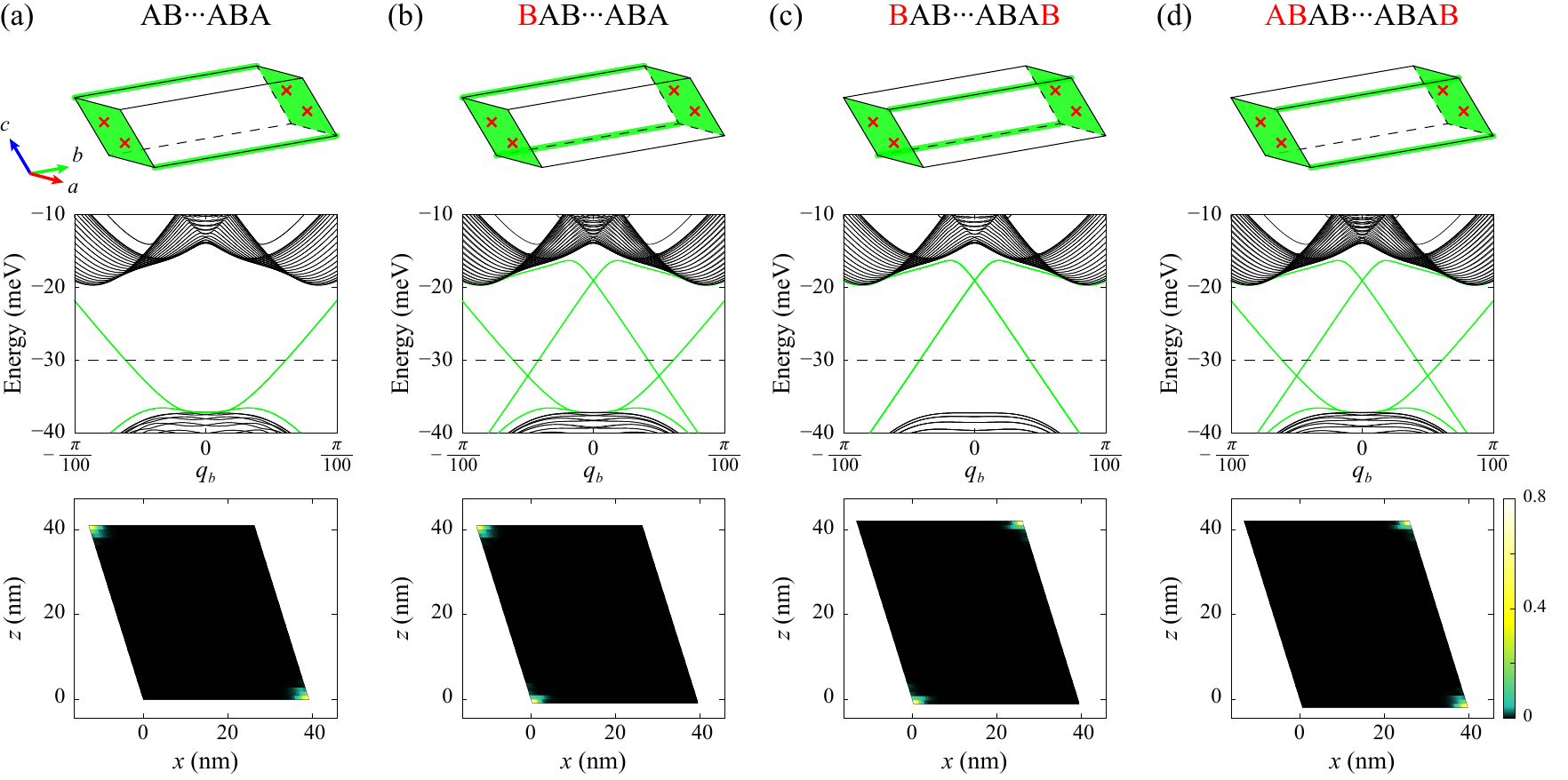}
\caption{The helical hinge states of $\alpha$-Bi$_{4}$Br$_{4}$ under four different scenarios  
of (001) surface terminations. (a)-(d) The label on the top of each column indicates 
the layer stacking order of the considered system in the $\bm c$ axis, 
and the red letters denote the extra (001) layer(s) compared with the case in (a). 
Upper panels: schematics of the helical hinge states in the $\bm b$ (chain) direction   
and the Dirac surface states at the (010) and $(0\bar{1}0)$ surfaces.
The green regions are the gapless boundaries, and each red cross denotes a 
surface Dirac cone. Middle panels: calculated band structures by using the MLWF. 
All the four systems are of finite size in the $\bm a$ and $\bm c$ directions 
and periodic in the $\bm b$ direction. The zero energy is set at the 0~meV in Fig.~\ref{fig6}(b).
The green lines are the helical hinge states, and the black lines are the gapped surface states. 
Because of the unbroken inversion symmetry in odd-layer systems, 
all the bands are doubly degenerate in (a) and (c). 
Lower panels: calculated spatial densities for the helical hinge states 
at the energy indicated by the dashed lines in the middle panels.}\label{fig1}
\end{figure*}

\subsection{Helical hinge states of $\alpha$-Bi$_{4}$Br$_{4}$}\label{sec:BiBr_hinge}

$\alpha$-Bi$_{4}$Br$_{4}$ has an AB stacking order in the (001) direction, 
similar to the Bernal stacked graphite. 
(The crystal structure of $\alpha$-Bi$_{4}$Br$_{4}$ is detailed in Sec.~\ref{crystal}.)  
Different from the Bernal graphite, 
the inversion center of $\alpha$-Bi$_{4}$Br$_{4}$ is not in the middle of a bilayer but in a monolayer, 
and each (001) monolayer of $\alpha$-Bi$_{4}$Br$_{4}$ is a 2D ${\mathbb Z}_2$ TI. 
It follows that there are four possible scenarios of (001) termination, i.e., A-A, B-A, B-B, and A-B,
and that they feature distinct patterns of helical hinge states, as depicted in the upper panels of Fig.~\ref{fig1}.  

In Fig.~\ref{fig1}(a), the featured system is periodic in the $\bm b$ direction, 
60-unit-cell long in the $\bm a$ direction, 
and 43-layer thick in the $\bm c$ direction with the A-A termination.
Because the system has an odd number of (001) layers, 
from the perspective of 2D ${\mathbb Z}_2$ TI, 
there must be one pair of helical edge states in total in the $\bm b$ direction. 
Given the unbroken inversion symmetry, the pair must be degenerate in energy 
and localized in two hinges that are reflected to each other under inversion.
The spectroscopic and spatial patterns of the pair of helical hinge states, 
calculated based on the MLWF, are presented in the middle and lower panels of Fig.~\ref{fig1}(a). 
 
The system in Fig.~\ref{fig1}(b) has one extra (001) layer stacked to 
the bottom of the system in Fig.~\ref{fig1}(a). As a result, 
one helical edge state of the extra TI layer annihilates the bottom hinge state in Fig.~\ref{fig1}(a), 
whereas the other creates a new hinge state at the opposite side of the bottom in Fig.~\ref{fig1}(b).
As the inversion symmetry is broken in this even-layer system, 
the pair of hinge states are not related under inversion. 
In Fig.~\ref{fig1}(c), one extra (001) layer is stacked to the top of the system in Fig.~\ref{fig1}(b). 
In a similar fashion, the top hinge state in Fig.~\ref{fig1}(b) is annihilated 
whereas a new hinge state emerges at the opposite side of the top in Fig.~\ref{fig1}(c). 
Since the inversion symmetry is restored in this odd-layer system, 
the new pair of hinge states become symmetric and degenerate. In Fig.~\ref{fig1}(d), 
the bottom of the system has one extra (001) layer compared to that in Fig.~\ref{fig1}(c). 
As expected, this extra TI layer breaks the inversion symmetry and 
switches the bottom hinge state to the opposite side. 

We note that, because of the ${\mathbb Z}_2$ character of 2D TI, 
the hinge states are gapped for the two even-layer scenarios  
yet remain gapless for the two odd-layer scenarios in the atomically thin limit. In the bulk limit, 
however, in each scenario the pair of hinge states are separate in space and gapless in energy.
While the scenarios in Figs.~\ref{fig1}(a) and~\ref{fig1}(c) are 
prototypes of the inversion symmetric time-reversal-invariant (TRI) TI with ${\mathbb Z}_4=2$~\cite{Tang2019a, Hsu2019}, 
here we demonstrate explicitly in Figs.~\ref{fig1}(b) and~\ref{fig1}(d) that 
the existence of helical hinge states does not require the inversion symmetry.
In fact, the hinge states in Figs.~\ref{fig1}(a) and~\ref{fig1}(c) are 
robust against inversion symmetry breaking, as long as the disturbance 
does not close the surface band gaps or hybridize the helical states at different hinges. 
Indeed, following the ZKM theory~\cite{Zhang2013}, the hinge states of $\alpha$-Bi$_{4}$Br$_{4}$ 
can also be demonstrated by applying a surface topological invariant 
or a surface domain-wall argument to an effective tight-binding model. 
The model and the two demonstrations are provided in Sec.~\ref{models} .

Evidently in Fig.~\ref{fig1}, depending on its (001) termination, 
each scenario exhibits a distinct pattern of helical hinge states.  
In fact, the four scenarios in Fig.~\ref{fig1} are the elementary building blocks 
of $\alpha$-Bi$_{4}$Br$_{4}$, 
and any many-layer system even with stacking faults can be decomposed into them. 
For instance, stacking Fig.~\ref{fig1}(c) on top of Fig.~\ref{fig1}(a) 
produces a scenario represented by Fig.~\ref{fig1}(d). 
This fact can be verified easily by superimposing the two schematics in the upper panels 
or the two band structures in the middle panels. 
Intriguingly, stacking Fig.~\ref{fig1}(c) on top of Fig.~\ref{fig1}(d) yields one layer of stacking fault: 
effectively, one 2D TI layer is embedded into the interior of Fig.~\ref{fig1}(d). 

\subsection{Helical hinge states of $\alpha$-Bi$_{4}$I$_{4}$}\label{sec:BiI_hinge}

$\alpha$-Bi$_{4}$I$_{4}$ is similar to $\alpha$-Bi$_{4}$Br$_{4}$ in two aspects. 
First, the primitive unit cell of $\alpha$-Bi$_{4}$I$_{4}$ also consists of two (001) layers. As a result, 
there are also four possible scenarios of (001) termination, i.e., A-B, B-B, B-A, and A-A.
Any many-layer system of $\alpha$-Bi$_{4}$I$_{4}$ even with stacking faults can be decomposed into them.
Second, each (001) monolayer of $\alpha$-Bi$_{4}$I$_{4}$ is a 2D ${\mathbb Z}_{2}$ TI 
with a crystal structure similar to that of $\alpha$-Bi$_{4}$Br$_{4}$ as detailed in Sec.~\ref{crystal}.

\begin{figure*}[t!]
\includegraphics[scale=1]{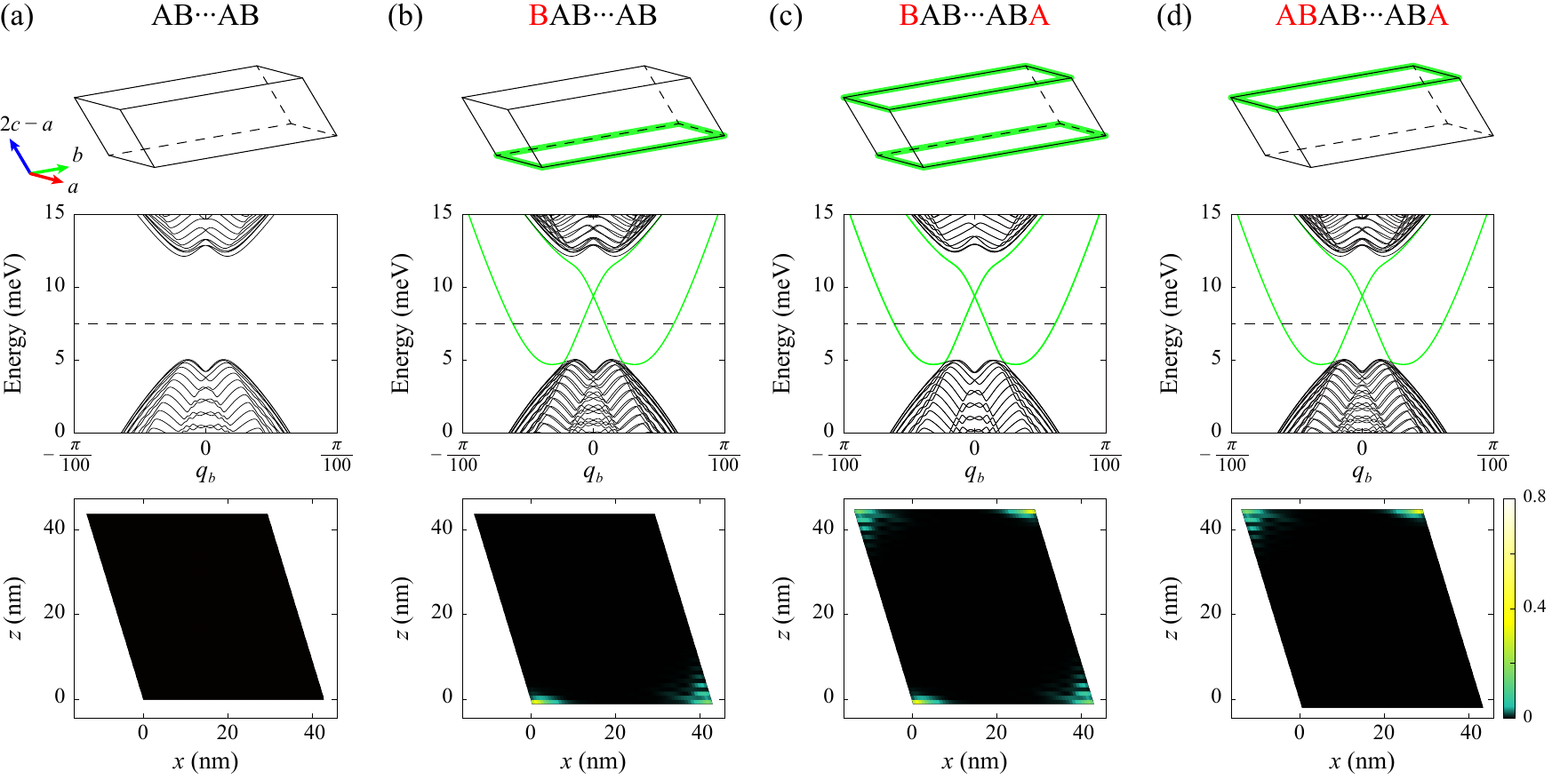}
\caption{The helical hinge states of $\alpha$-Bi$_{4}$I$_{4}$ under four different scenarios  
of (001) surface terminations. (a)-(d) The label on the top of each column indicates 
the layer stacking order of the considered system in the $2{\bm c}-{\bm a}$ axis, 
and the red letters denote the extra (001) layer(s) compared with the case in (a). 
Upper panels: schematics of the helical hinge states in the $\bm b$ (chain) direction.
The green regions are the gapless boundaries. Middle panels: calculated band structures by using the MLWF. 
All the four systems are of finite size in the $\bm a$ and $2{\bm c}-{\bm a}$ directions 
and periodic in the $\bm b$ direction. The zero energy is set at the -8~meV in Fig.~\ref{fig6}(c).
The green lines are the helical hinge states, and the black lines are the gapped surface states. 
Because of the unbroken inversion symmetry in even-layer systems, 
all the bands are doubly degenerate in (a) and (c). 
Lower panels: calculated spatial densities for the helical hinge states 
at the energy indicated by the dashed lines in the middle panels.}\label{fig2}
\end{figure*}

However, in contrast to the case of $\alpha$-Bi$_{4}$Br$_{4}$, 
the inversion center of $\alpha$-Bi$_{4}$I$_{4}$ is in the middle of two adjacent (001) layers.
Since the A- and B-type layers are reflected to each other under inversion, 
only even-layer systems are inversion symmetric. 
Significantly, although the bulk $\alpha$-Bi$_{4}$I$_{4}$ has ${\mathbb Z}_4=0$ 
as an inversion symmetric TRI insulator and trivial symmetry indicators in general~\cite{Zhang2019, Vergniory2019, Tang2019},
we explicitly show in Fig.~\ref{fig2} that $\alpha$-Bi$_{4}$I$_{4}$ 
can host helical hinge states for three of its four possible (001) terminations.

In Fig.~\ref{fig2}(a), the featured system is periodic in the $\bm b$ direction, 
60-unit-cell long in the $\bm a$ direction, 
and 44-layer thick in the $2{\bm c}-{\bm a}$ direction with the A-B termination.
The spectroscopic and spatial patterns of this system, calculated based on the MLWF, 
are presented in the middle and lower panels of Fig.~\ref{fig2}(a). 
Clearly, this system of A-B termination does not host any gapless boundary state.
This result is consistent with the fact that the system in Fig.~\ref{fig2}(a) has 
an even number of 2D ${\mathbb Z}_{2}$ TIs from the 2D perspective and the fact that 
$\alpha$-Bi$_{4}$I$_{4}$ has trivial symmetry indicators from the 3D perspective~\cite{Zhang2019, Vergniory2019, Tang2019}. 
The system of B-B termination in Fig.~\ref{fig2}(b) has one extra (001) layer stacked to 
the bottom of the system in Fig.~\ref{fig2}(a). 
The new system can be viewed as a 2D ${\mathbb Z}_{2}$ TI 
since it consists of an odd number of (001) layers.
As a result, a pair of helical hinge states in the $\bm b$ direction 
emerges at the opposite sides of the bottom in Fig.~\ref{fig2}(b).
For the B-A termination in Fig.~\ref{fig2}(c), 
one extra (001) layer is stacked to the top of the system in Fig.~\ref{fig2}(b). 
In a similar fashion, a pair of helical hinge states emerges 
at the opposite sides of the top in Fig.~\ref{fig2}(c), in addition to the pair at the bottom.
This even-layer system is trivial from the perspectives of 2D ${\mathbb Z}_{2}$ TI and 3D symmetry indictors, 
yet it hosts helical hinge states at all the four hinges in the $\bm b$ direction.
For the A-A termination in Fig.~\ref{fig2}(d), 
the bottom of the system has one extra (001) layer compared to that in Fig.~\ref{fig2}(c). 
As expected, the edge states of the extra TI layer annihilate the bottom hinge states in Fig.~\ref{fig2}(c). 
 
We note that the two odd-layer scenarios in Figs.~\ref{fig2}(b) and~\ref{fig2}(d) have neither 
the inversion symmetry nor the $C_{2}$ symmetry of the bulk as detailed in Sec.~\ref{crystal}. 
In each scenario the pair of hinge states is not related by any symmetry. 
Intriguingly, the two scenario can be related under inversion,  
and thus all the features in Figs.~\ref{fig2}(b) and~\ref{fig2}(d) are the same,   
except that their hinge states are localized at the opposite hinges.
Nevertheless, the three patterns of hinge states in Fig.~\ref{fig2} are all  
robust against inversion symmetry breaking, as long as the disturbance does not close 
the surface band gaps or hybridize the helical states at different hinges.  
We stress that, whereas the hinge states of $\alpha$-Bi$_{4}$I$_{4}$ 
cannot be understood by the bulk symmetry indicators, 
they can be captured within the ZKM theory~\cite{Zhang2013} by applying a surface topological invariant 
or a surface domain-wall argument to an effective tight-binding model, as demonstrated in Sec.~\ref{models}.

\subsection{Surface SSH model: coupled edge construction}\label{sec:SSH}

The key features of $\alpha$-Bi$_{4}$Br$_{4}$ and $\alpha$-Bi$_{4}$I$_{4}$  
presented in Figs.~\ref{fig1} and~\ref{fig2} are all calculated by using the DFT-based MLWF. 
In Secs.~\ref{sec:BiBr_hinge} and~\ref{sec:BiI_hinge} we also deduce all these features based on 
the minimal numerical results, i.e., one of the four scenarios, and the knowledge of 2D ${\mathbb Z}_{2}$ TI.
To further explain these appealing features, 
now we provide a computation-free surface argument 
based on {the locations of inversion centers} and the knowledge of 2D ${\mathbb Z}_{2}$ TI.
(A similar argument also exists by considering the locations of twofold rotation axes.)
Because each (001) monolayer is a 2D ${\mathbb Z}_{2}$ TI, 
the (100) side surface of $\beta$-Bi$_{4}$X$_{4}$ or $\alpha$-Bi$_{4}$Br$_{4}$ 
can be viewed as a ``chain'' (in the $\bm c$ direction) of coupled helical edge states 
(in the $\bm b$ direction), and likewise the ($\bar{1}$00) side surface. 
This fact also applies to the $(201)$ and $(\bar{2}0\bar{1})$ side surfaces of $\alpha$-Bi$_{4}$I$_{4}$.
(The subtle differences between the crystal structures of the two $\alpha$ phases are detailed in Sec.~\ref{crystal}).
This argument is analogous to the Su-Schrieffer-Heeger (SSH) model~\cite{Su1979}, 
and it is valid because the nearest-neighbor (NN) inter-layer edge tunnelings are much smaller than the bulk band gaps. 

\begin{figure}[t!]
\includegraphics[scale=1]{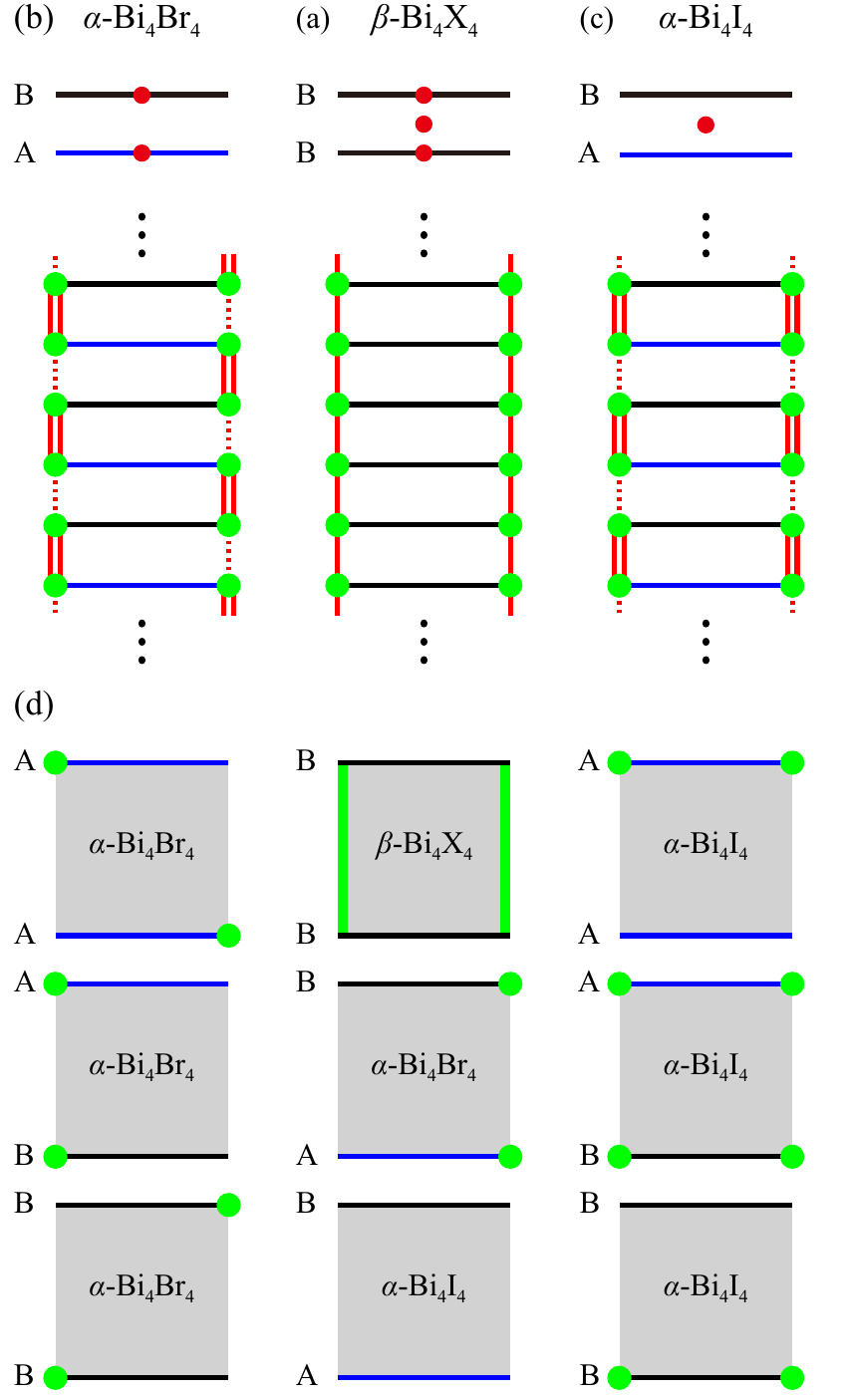}
\caption{(a)-(c) The possible inversion centers (or twofold rotation axes) of 
Bi$_{4}$X$_{4}$ and the dimerization patterns of the coupled edge states at the two side surfaces. 
The black and blue lines denote the B- and A-type (001) layers.
The red dots denote the inversion centers (or twofold rotation axes).
The green dots denote the helical edge states in the $\bm b$ (chain) direction.
The double solid and single dashed red lines denote the stronger and weaker NN inter-edge tunnelings. 
(d) All possible (001) surface terminations: one for the $\beta$ phase and four for each $\alpha$ phase. 
The green lines denote the Dirac surface states, and the green dots denote the helical hinge states.} 
\label{fig3}
\end{figure}

Consider first the WTI $\beta$-Bi$_{4}$X$_{4}$~\cite{Liu2016, Noguchi2019} depicted in Fig.~\ref{fig3}(a).  
The primitive unit cell of $\beta$-Bi$_{4}$X$_{4}$ consists of only one (001) layer. 
The inversion center can be placed in a monolayer or in the middle of a bilayer. 
As a result, the inversion symmetry restricts 
the NN inter-edge tunnelings to be the same between any two adjacent edges at any side surface. 
Given that the tunnelings are weak, each side surface states 
can be viewed as a 1D Dirac cone along $k_b$ dispersing weakly along $k_c$; 
the Dirac cone is gapless only at the TRI points $k_c=0$~and~$\pi$. 
This is exactly what has been predicted in a previous theory~\cite{Liu2016} 
and observed in a recent experiment~\cite{Noguchi2019}.
Therefore, both the (100) and ($\bar{1}$00) side surfaces are gapless,  
independent of the number of (001) layers, as sketched in Fig.~\ref{fig3}(d).
This is analogous to the critical point of the SSH model. Remarkably, 
the $\beta$ phase (an equally spaced chain of edge states) does undergo a Peierls transition 
(i.e., dimerization) to an $\alpha$ phase (dimerized chain of edge states) at low temperature~\cite{Liu2016}. 
The critical temperature of Bi$_{4}$I$_{4}$ turns out to be room temperature~\cite{VonSchnering1978, Dikarev2001, Weiz2017, Noguchi2019, Note101}. 
There may exist an edge-state Peierls' theorem to explain the instability.
% Peierls' theorem states that a one-dimensional equally spaced chain with one electron per ion is unstable. 

In the case of $\alpha$-Bi$_{4}$Br$_{4}$, the inversion center 
can only be placed in a (001) layer. Consequently, as depicted in Fig.~\ref{fig3}(b), 
the inversion symmetry only relates the NN upper A-lower B tunneling 
at one side surface to the NN lower A-upper B tunneling at the other side surface, 
and the NN upper A-lower B and lower A-upper B tunnelings at the same side surface 
generally have different strengths. 
This gives rise to the unique dimerization pattern in Fig.~\ref{fig3}(b): 
the two side surfaces exhibit opposite dimerizations.
Following the spirit of SSH model, the dimerizations gap the strongly coupled edge states, 
and any weakly coupled one left by a (001) surface termination yields a helical hinge state.
As showcased in Fig.~\ref{fig3}(d), there are four possible terminations,  
and each has a distinct pattern of hinge states. 
This explains the four scenarios of hinge states featured in Fig.~\ref{fig1}. 
  
In the case of $\alpha$-Bi$_{4}$I$_{4}$, by contrast, 
the inversion center can only be placed between two adjacent layers. 
Accordingly, as depicted in Fig.~\ref{fig3}(c), 
the NN upper A-lower B tunnelings at different side surfaces are related by the inversion symmetry, 
and likewise the NN lower A-upper B tunnelings. 
Yet, the two tunnelings generally have different strengths.
It follows that the two side surfaces exhibit the same dimerization. 
Moreover, the strongly coupled edge states become gapped, 
leaving the weakly coupled ones at the hinges gapless. 
As showcased in Fig.~\ref{fig3}(d), there are four possible terminations,  
and three yield distinct hinge states while one is trivial.
This explains the four scenarios of hinge states featured in Fig.~\ref{fig2}. 

Clearly, a key difference between the two $\alpha$ phases is whether 
the two side surfaces exhibit the same or opposite dimerizations. 
For $\alpha$-Bi$_{4}$Br$_{4}$, because of the opposite dimerizations at the two side surfaces, 
there always exists two unpaired edge states, one at the top and the other at the bottom, 
independent of the termination. This is rooted in the fact that 
its inversion center can only be placed in a (001) layer. 
For $\alpha$-Bi$_{4}$I$_{4}$, because of the same dimerization at the two side surfaces, 
the unpaired edge states always appear in pair at the top or bottom (or both). This originates from the fact that 
its inversion center can only be placed between two adjacent layers. 
We point out that compelling evidence of the revealed surface dimerization patterns 
in the two $\alpha$ phases is provided by their crystal structures in Fig.~\ref{fig5} 
and effective surface models in Eqs.~(\ref{H-s-alpha1}) and~(\ref{H-s-alpha2}). 

\subsection{Rotational symmetry-protected surface states}\label{surface}

\begin{figure}[t!]
\includegraphics[scale=1]{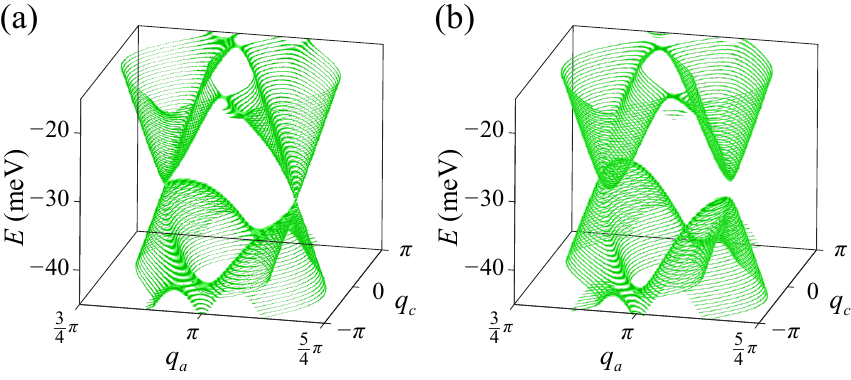}
\caption{(a) Gapless (010) surface states of pristine $\alpha$-Bi$_{4}$Br$_{4}$ 
and (b) gapped (010) surface states of $\alpha$-Bi$_{4}$Br$_{4}$ with 
a $C_{2}$-asymmetric TRI surface potential. 
The constant-energy contours with a spacing of $0.5$~meV are obtained 
by the surface Green's function calculations of the MLWF for a semi-infinite system. 
The zero energy is set at the 0~meV in Fig.~\ref{fig6}(b).}
\label{fig4} 
\end{figure}

For completeness, we close this section by showing that $\alpha$-Bi$_{4}$Br$_{4}$ also hosts 
protected surface states at the (010) and (0$\bar 1$0) surfaces. 
The gapless surface Dirac cones are denoted as the red crosses in Fig.~\ref{fig1}. 
Based on the surface Green's function calculations of the MLWF for a semi-infinite system,  
Fig.~\ref{fig4}(a) features the two $(010)$ surface Dirac cones of $\alpha$-Bi$_{4}$Br$_{4}$, 
which are protected and related by the $\mathcal{C}_2$ symmetry. 
By contrast, a similar calculation for $\alpha$-Bi$_{4}$I$_{4}$ reveals no gapless surface states, 
although it also has the $\mathcal{C}_2$ symmetry. 
Nevertheless, we confirm the previous finding~\cite{Tang2019,Hsu2019} of 
$\alpha$-Bi$_{4}$Br$_{4}$ being a rare 
topological crystalline insulator with a surface rotation anomaly~\cite{Fang2019} 
by using more accurate first-principles calculations as detailed in Sec.~\ref{sec:band}. 
One can further verify this result directly by evaluating a rotation invariant~\cite{Fang2019} 
or indirectly by using the symmetry indicators~\cite{Tang2019,Hsu2019} 
along with the inversion eigenvalues listed in Sec.~\ref{sec:band}.

Moreover, when the effects such as dangling bonds and surface reconstruction are ignored,  
the two surface Dirac points are identified at $(q_{a}, q_{c})=\pm(0.861, 0.113)\pi$. 
When a TRI (010) surface potential that breaks the $\mathcal{C}_2$ symmetry is added in our calculation,
the gapless Dirac cones in Fig.~\ref{fig4}(a) become gapped, as shown in Fig.~\ref{fig4}(b). 
This unambiguously demonstrates that it is the $\mathcal{C}_2$ symmetry 
that protects the (010) surface Dirac cones.
We point out that it would be challenging to observe the gapless (010) surface states in experiment, 
because the (010) surface is not a natural cleavage plane, and because $\alpha$-Bi$_{4}$Br$_{4}$ is extremely soft~\cite{Note101}. 
Most likely, the $\mathcal{C}_2$ symmetry would be broken by the (010) dangling bonds and surface reconstruction.
Fortunately, the hinges states of our major interest are between two natural cleavage planes.
This facilitates the future hinge state experiments. 

\section{Crystal Structures}\label{crystal}

Both the $\alpha$ and $\beta$ phases of Bi$_{4}$X$_{4}$ crystallize in the same monoclinic space group 
$C^{3}_{2h}$ ($C2/m$). They have three spatial symmetries: 
inversion ($\mathcal{P}$), (010) mirror reflection ($\mathcal{M}_b$), 
and twofold rotation around the $\bm b$ axis ($\mathcal{C}_2$). 
Given that $\mathcal{C}_{2}=\mathcal{M}_{b}\mathcal{P}$, only two of the three symmetries are independent. 
The building block of Bi$_{4}$X$_{4}$ is an atomic chain with strongly covalent bonds between bismuth atoms. 
Each chain consists of four inequivalent Bi atoms and four inequivalent X atoms. As shown in Fig.~\ref{fig5}(a), 
the two internal (external) Bi atoms are denoted by Bi$_{\rm in}$ and Bi$_{\rm in}'$ (Bi$_{\rm ex}$ and Bi$_{\rm ex}'$), 
and the angles between different Bi-Bi bonds are denoted by four $\theta$'s. 
The chains are oriented in the $\bm b$ direction and stacked 
via the van der Waals forces in the $\bm a$ and $\bm c$ directions~\cite{VonBenda1978, VonSchnering1978, Liu2016}.

We note that Bi$_{4}$X$_{4}$ has two natural cleavage surfaces. For $\beta$-Bi$_{4}$Br$_{4}$, 
the inter-layer binding energy for the (001) and (100) planes are respectively 20 and 25 meV/$\rm \AA^{2}$ 
(slightly larger than that of graphite and smaller than that of MoS$_2$)~\cite{Liu2016}.  
These values are consistent with the fact that for the inter-chain distances 
$c$ is slightly larger than $a/2$ as detailed in Appendix~\ref{data}. 
For the other three materials, the binding energies are in the same range.
Nevertheless, this unique property of quasi-1D materials 
highlights the $\beta$ phase as a prototype WTI~\cite{Liu2016} that bears surface selective hallmarks. 
Moreover, the absence of dangling bonds for hinges between the two cleavage surfaces 
renders the two $\alpha$ phases ideal platforms for exploring HOTIs that host helical hinge states.

\subsection{Crystal structure of $\beta$-Bi$_{4}$X$_{4}$} 

While $\beta$-Bi$_{4}$I$_{4}$ has been experimentally synthesized~\cite{VonSchnering1978, Autes2016, Noguchi2019}, 
$\beta$-Bi$_{4}$Br$_{4}$ is a designed material~\cite{Liu2016} 
based on the same crystal structure of $\beta$-Bi$_{4}$I$_{4}$. 
The dynamic stability of $\beta$-Bi$_{4}$Br$_{4}$ has been 
demonstrated through the phonon spectrum calculations~\cite{Liu2016}. 
Nevertheless, their detailed crystal structure data are provided in Appendix~\ref{data}.

Each atomic chain is inversion symmetric in $\beta$-Bi$_{4}$X$_{4}$, as seen in Fig.~\ref{fig5}(a).
Under inversion, the Bi$_{\rm in}$ and Bi$_{\rm ex}$ atoms are reflected into 
the Bi$_{\rm in}'$ and Bi$_{\rm ex}'$ atoms, respectively. 
In addition, each chain is mirror-symmetric with respect to any (010) plane that contains the Bi atoms.
The mirror and inversion symmetries dictate that $\theta_{1} = \theta_{1}'$ and $\theta_{2} = \theta_{2}'$, 
respectively. 

Each (001) monolayer of $\beta$-Bi$_{4}$X$_{4}$ consists of equally spaced atomic chains 
in the $\bm a$ direction, as shown in Fig.~\ref{fig5}(b).
Two adjacent chains are displaced from each other by constant vectors $a_{1,2}=({\bm a}\mp{\bm b})/2$.
The monolayer is inversion- and mirror-symmetric, 
with the same inversion centers and mirror planes as its individual chains. 
In addition to the intra-chain inversion centers off the mirror planes, 
there exist inter-chain inversion centers in the mirror planes between two adjacent chains, 
as shown in Fig.~\ref{fig5}(b).  
It follows that the monolayer is also invariant under a twofold rotation ($\mathcal{C}_{2}=\mathcal{M}_{b}\mathcal{P}$) 
around the $\bm b$ axis across the inter-chain inversion centers. 

As shown in Fig.~\ref{fig5}(c), bulk $\beta$-Bi$_{4}$X$_{4}$ is a periodic stack of (001) layers 
in the $\bm c$ axis, which is normal to the $\bm b$ axis and 
$107.87^{\circ}$ ($\beta$ in Appendix~\ref{data}) above the $\bm a$ axis. 
Given that $\bm c \perp \bm b$, the bulk crystal has the same mirror planes as its individual monolayers.  
Remarkably, the bulk inversion center of $\beta$-Bi$_{4}$X$_{4}$ can be placed 
not only in a (001) layer but also in the middle of two adjacent layers, 
and likewise the twofold rotation axis. 

\subsection{Crystal structure of $\alpha$-Bi$_{4}$Br$_{4}$}

\begin{figure}[p]
\includegraphics[scale=1]{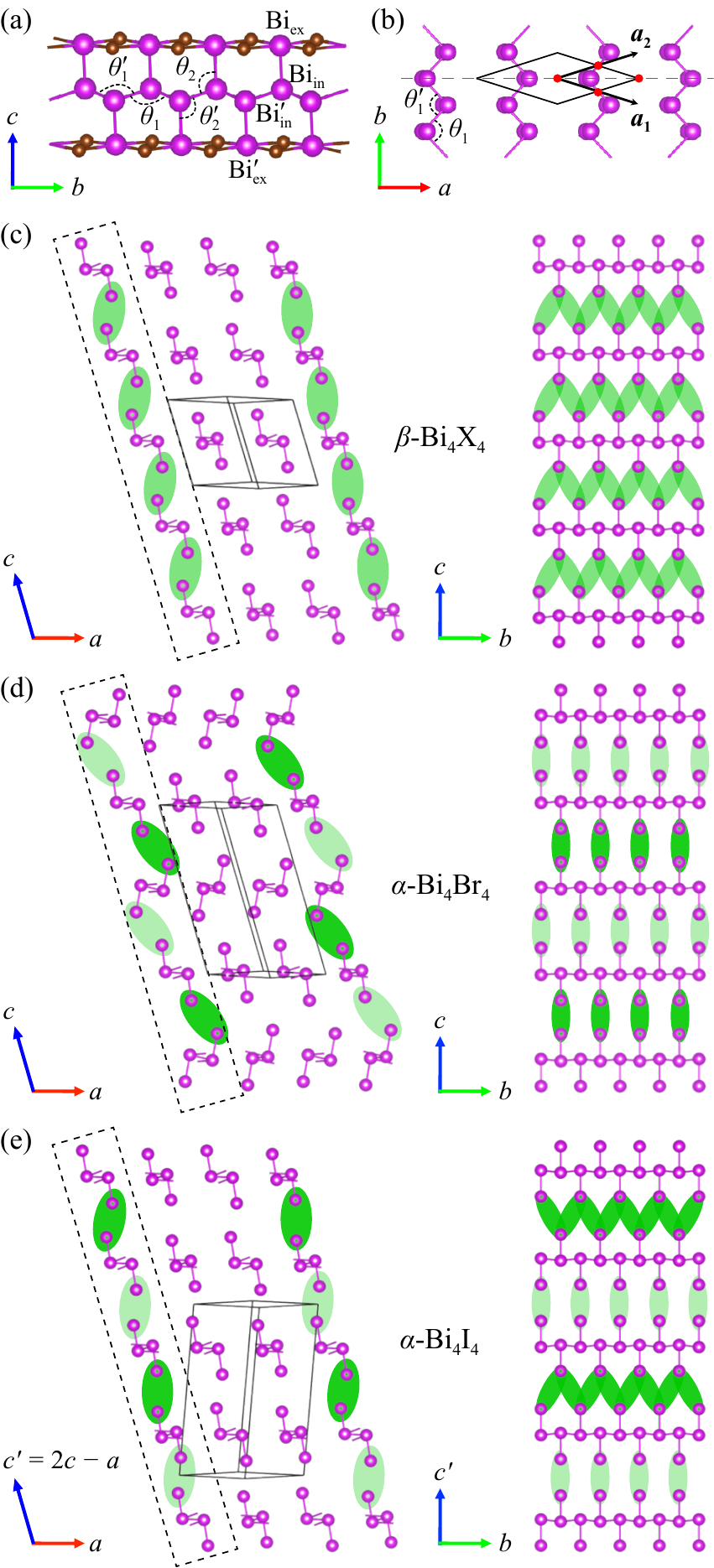}
\caption{(a) An atomic chain as the building block of Bi$_{4}$X$_{4}$. 
The purple (brown) balls are the Bi (X) atoms. (b) A (001) monolayer of Bi$_{4}$X$_{4}$. 
The black diamond is a primitive unit cell. The red dots are possible inversion centers. The dashed red line is a mirror plane. 
(c)-(e) Left panels: the bulk structures of $\beta$-Bi$_{4}$X$_{4}$, 
$\alpha$-Bi$_{4}$Br$_{4}$, and $\alpha$-Bi$_{4}$I$_{4}$ viewed from the $\bar{\bm b}$ axis. 
The solid black lines sketch the primitive unit cells. 
The X atoms are omitted for better illustration. 
The green bubbles indicate the dimerization patterns of the NN inter-edge tunnelings.  
Right panels: the side monolayers in the dashed black frames in the left panels viewed from the $\bm a$ axis.} 
\label{fig5}
\end{figure}

Each (001) monolayer of $\alpha$-Bi$_{4}$Br$_{4}$ has the same crystal structure as that of 
$\beta$-Bi$_{4}$X$_{4}$, except for the slightly different lattice constants and intra-chain parameters 
listed in Appendix~\ref{data}. Unlike $\beta$-Bi$_{4}$X$_{4}$, 
two adjacent layers are not related by any symmetry, as shown in Fig.~\ref{fig5}(d). 
For instance, although the two layers have the same inter-chain distance in the $\bm a$ axis, 
the key intra-chain parameters $\theta$'s are different for the two layers. 
As a result, the primitive unit cell consists of two (001) layers. 
Bulk $\alpha$-Bi$_{4}$Br$_{4}$ has the same symmetries as $\beta$-Bi$_{4}$X$_{4}$:  
inversion, (010) mirror reflection, and twofold rotation around the $\bm b$ axis. However, 
the inversion center and the rotational axis of bulk $\alpha$-Bi$_{4}$Br$_{4}$ can only be placed in a (001) layer. 
The detailed crystal structure data are provided in Appendix~\ref{data}.

\subsection{Crystal structure of $\alpha$-Bi$_{4}$I$_{4}$}

For each atomic chain of $\alpha$-Bi$_{4}$I$_{4}$, 
the (010) mirror symmetry is preserved such that $\theta_{1} = \theta_{1}'$, 
whereas the inversion symmetry is broken as indicated by $\theta_{2}\neq\theta_{2}'$. 
Even for a (001) monolayer of $\alpha$-Bi$_{4}$I$_{4}$, only the mirror symmetry is present. 
For bulk $\alpha$-Bi$_{4}$I$_{4}$, however, the inversion and twofold rotational symmetries are restored, 
as each symmetry relates two adjacent layers. Similar to $\alpha$-Bi$_{4}$Br$_{4}$, 
the primitive unit cell of $\alpha$-Bi$_{4}$I$_{4}$ consists of two (001) layers. 
Different from $\alpha$-Bi$_{4}$Br$_{4}$, the inversion center and the rotational axis 
of bulk $\alpha$-Bi$_{4}$I$_{4}$ can only be placed in the middle of two adjacent layers. 
 
\begin{figure*}[t!]
\includegraphics[scale=1]{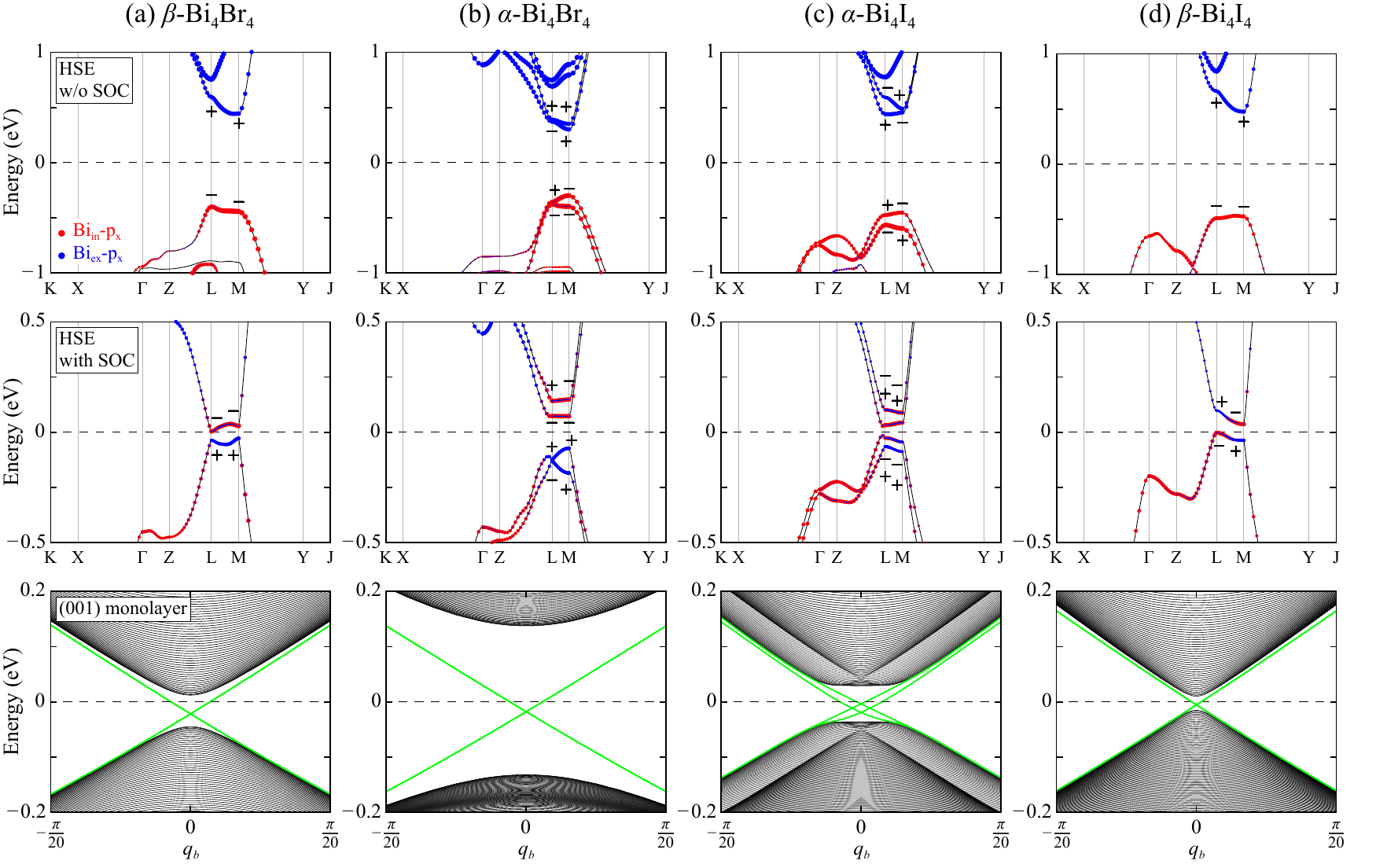}
\caption{The 3D bulk and (001) monolayer ribbon band structures of (a) $\beta$-Bi$_4$Br$_4$, 
(b) $\alpha$-Bi$_4$Br$_4$, (c) $\alpha$-Bi$_4$I$_4$, and (d) $\beta$-Bi$_4$I$_4$.
The SOC is not included in the top panels but included in the middle and bottom panels. 
The size of red (blue) dots indicates the weight of the $p_x$ orbital of $\rm Bi_{in}$ ($\rm Bi_{ex}$) atoms.  
The $\pm$ signs label the inversion eigenvalues at the $L$ and $M$ points, 
with the inversion centers in (001) layers in (a), (b), and (d) but in the middle of two adjacent layers in (c).
The green lines are the helical edge states (degenerate for two edges in (a), (b), and (d)),  
implying that the four (001) monolayers are all 2D ${\mathbb Z}_2$ TIs.}\label{fig6}
\end{figure*} 
 
Bulk $\alpha$-Bi$_{4}$I$_{4}$ exhibits a clear difference from the other three materials 
in how the (001) layers are stacked, as shown in Fig.~\ref{fig5}(e). 
This is best indicated by the angle between the $\bm c$ and $\bm a$ axes, $\beta = 87.04^{\circ}$~\footnotetext[102]{
Note that $\beta$ can be either $87.04^{\circ}$ or $92.96^{\circ}$ for $\alpha$-Bi$_4$I$_4$,   
and that conventionally the obtuse angle is chosen. Here we choose $\beta=87.04^{\circ}$ instead for the consistency with the coordinates of the other three materials.}\cite{Note102},   
which is significantly different from $\sim107^{\circ}$ of the other three materials. 
If ${\bm c'} = 2{\bm c} - {\bm a}$ was the primitive lattice vector, 
the stacking direction would be the same as the other three materials, 
but there would be four layers per primitive unit cell. 
Nevertheless, there exists a shorter inter-layer lattice vector $\bm c$,  
and the true primitive unit cell only contains two layers. 
As such, the natural cleavage surface of $\alpha$-Bi$_{4}$I$_{4}$, 
corresponding to the (100) surface of the other three materials, is the (201) surface~\cite{VonSchnering1978}. 
The detailed crystal structure data are provided in Appendix~\ref{data}.

\subsection{Atomic dimerization in $\alpha$-Bi$_{4}$X$_{4}$}\label{ADim}

From the crystal structures in Fig.~\ref{fig5}, 
the dimerization patterns at the side surfaces sketched in Fig.~\ref{fig3} can be visualized clearly. 
For $\beta$-Bi$_{4}$X$_{4}$ in Fig.~\ref{fig5}(c), the NN inter-edge tunnelings are the same 
at both the (100) and ($\bar 1$00) side surfaces between any two adjacent (001) layers. Thus, 
the two surfaces share the same dimerization pattern resembling the critical point of the SSH model. 
For $\alpha$-Bi$_{4}$Br$_{4}$ in Fig.~\ref{fig5}(d), the highlighted 
NN inter-layer tunnelings are the same in the bulk between any two adjacent (001) layers. 
At the (100) and ($\bar 1$00) side surfaces, however, 
the closer the Bi$_{\rm ex}$/Bi$_{\rm ex}'$ atoms are to the vacuum, 
the weaker their tunnelings are. 
This leads to opposite dimerization patterns at the two side surfaces. 
By sharp contrast, for $\alpha$-Bi$_{4}$I$_{4}$ in Fig.~\ref{fig5}(e), the highlighted 
NN inter-layer tunnelings are even evidently different in the bulk. 
It follows that the dimerization patterns at the (201) and ($\bar2$0$\bar1$) side surfaces are the same.

\section{Band Structures}\label{sec:band}

With the crystal structures of Bi$_{4}$X$_{4}$ in Fig.~\ref{fig5} and Appendix~\ref{data}, 
we carry out the DFT calculations to obtain their electronic band structures and analyze their topological band properties.
The DFT calculations were performed by using the projector augmented wave method 
implemented in the Vienna {\it ab initio} simulation package~\cite{Kresse1996} 
and the Perdew-Burke-Ernzerhof parametrization of the generalized gradient approximation 
for the exchange correlation potential~\cite{Perdew1996, Kresse1999}. 
In order to obtain more accurate band gaps and band inversions,  
we apply the more sophisticated Heyd-Scuseria-Ernzerhof (HSE) 
hybrid functional method~\cite{Heyd2003} to the calculations. 
We employ the DFT results and the Wannier90 code~\cite{Marzari1997, Souza2001, Mostofi2008} 
to construct the MLWF for the $p$ orbitals of Bi and halogens. Based on the MLWF, 
we derive the electronic band structures in Figs.~\ref{fig1}-\ref{fig3} for finite-size systems. 
The methods here are the same as those in our previous work~\cite{Liu2016}.  
Figure~\ref{fig6} displays the bulk band structures and band inversions for 
$\beta$-Bi$_4$Br$_4$, $\alpha$-Bi$_4$Br$_4$, $\alpha$-Bi$_4$I$_4$, and $\beta$-Bi$_4$I$_4$.

\subsection{Symmetry indicators of Bi$_{4}$X$_{4}$} 

Informed by the band inversions in Fig.~\ref{fig6}, 
we can obtain the symmetry indicators of the space group $C2/m$ (No.~12) 
$({\mathbb Z}_2, {\mathbb Z}_2, {\mathbb Z}_2; {\mathbb Z}_4)$~\cite{Po2017, Khalaf2018a}: 
$(001;2)$ for $\beta$-Bi$_4$Br$_4$, % (001;0)
$(110;1)$ for $\beta$-Bi$_4$I$_4$,  % (110;3)
and $(000;2)$ for $\alpha$-Bi$_4$Br$_4$ with their inversion centers placed in (001) monolayers 
and $(000;0)$ for $\alpha$-Bi$_4$I$_4$ with its inversion center placed in the middle of two adjacent (001) layers. 
The first three ${\mathbb Z}_2$ indices are the Fu-Kane weak indices~\cite{Fu2007}, 
and the ${\mathbb Z}_4$ index is the total number of band inversions modulo 4~\cite{Po2017, Khalaf2018a}. 
$\beta$-Bi$_4$Br$_4$ is a prototype WTI that can be viewed as 
a periodic stack of 2D ${\mathbb Z}_2$ TIs with one TI layer per unit cell.  
While our calculation predicts $\beta$-Bi$_4$I$_4$ to be a STI, 
two recent ARPES experiments obtained contrasting conclusions: STI versus WTI~\cite{Autes2016, Noguchi2019}.  
In fact, a small strain can tune $\beta$-Bi$_4$I$_4$ to a WTI~\cite{Liu2016}, 
and the (001) monolayer of $\beta$-Bi$_4$I$_4$ is indeed a 2D ${\mathbb Z}_2$ TI as shown in Fig.~\ref{fig6}.
For the purpose of understanding the two $\alpha$ phases, 
we view the two $\beta$ phases as the WTI with interlayer couplings much smaller than band gaps.
Notably, in the same classification based on the symmetry indicators, 
while $\alpha$-Bi$_4$Br$_4$ is a HOTI with ${\mathbb Z}_4=2$, 
$\alpha$-Bi$_4$I$_4$ is topologically trivial in all possible classes. 
However, $\alpha$-Bi$_4$I$_4$ is also a true HOTI as clearly evidenced in Fig.~\ref{fig2}. 

\subsection{Unit cell doubling of $\beta$-Bi$_{4}$X$_{4}$}\label{doubling}

Now we show that the band inversions and symmetry indicators of the two $\alpha$ phases can be directly 
understood by applying zone folding to the $\beta$ phase WTI yet choosing two different locations for their inversion centers. 

Doubling the unit cell in the $\bm c$ axis folds the TRI momenta with $q_{3} = \pi$ back to those with $q_{3} = 0$. 
This implies that, due to the zone folding, all the band inversions in the $q_{3} = \pi$ plane 
move to the corresponding TRI momenta in the $q_{3} = 0$ plane.
Moreover, at the new TRI momenta in the reduced Brillouin zone (BZ) with $q_{3}' = \pi$,
the inversion eigenstates are symmetric and antisymmetric combinations 
of the band states at $q_{3} = \pm \pi /2$ in the original BZ, i.e., 
$\ket{q_{3}' = \pi, \pm} = \left(\ket{q_{3} = {\pi}/{2}} \pm \ket{q_{3} = -{\pi}/{2}}\right)/\sqrt{2}$, 
where $\mathcal{P}\ket{q_{3} = \pm{\pi}/{2}} =\ket{q_{3} = \mp{\pi}/{2}}$ 
and $q_1,\,q_2=0$ or $\pi$ implicitly. 
Consequently, the band states at the TRI momenta with $q_{3}' = \pi$ are four-fold degenerate, 
and for each degeneracy the two Kramers pairs have opposite inversion eigenvalues. 
This implies that the TRI momenta with $q_{3}' = \pi$ are irrelevant 
to the ${\mathbb Z}_2$ and ${\mathbb Z}_4$ indices. 

\begin{figure}[t!]
\includegraphics[scale=1]{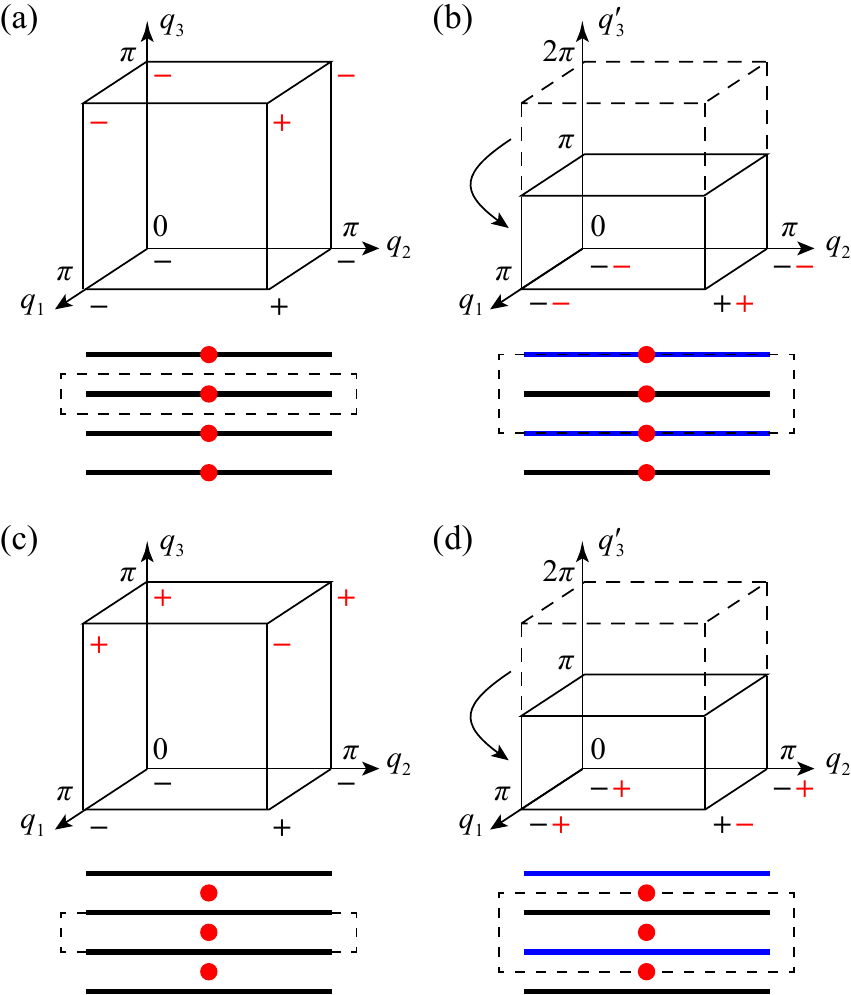}
\caption{The inversion eigenvalues of the occupied bands at the TRI momenta 
for the $\beta$ phase WTI, 
modeled by Eqs.~(\ref{H-beta}) and~(\ref{H-beta2}). 
Different inversion centers and primitive unit cells are chosen in (a) and (c). 
Illustrating the band inversions in $\alpha$-Bi$_4$Br$_4$ and $\alpha$-Bi$_4$I$_4$, respectively, 
(b) and (d) are the doubled unit cell counterparts of (a) and (c). 
In the bottom panels, the solid lines, dashed boxes, and red dots denote the (001) layers, 
unit cells, and inversion centers, respectively.}
\label{fig7}
\end{figure}

When the inversion center is placed in a (001) monolayer, 
Fig.~\ref{fig7}(a) sketches the two band inversions of the $\beta$ phase WTI 
at $(\pi,\pi,0)$ and $(\pi,\pi,\pi)$ in the original BZ. This implies the symmetry indicators $(001;2)$.
For the same choice of inversion center, the two band inversions appear at $(\pi,\pi,0)$ 
in the reduced BZ as sketched in Fig.~\ref{fig7}(b),
and the symmetry indicators become $(000;2)$, which also characterizes $\alpha$-Bi$_4$Br$_4$. 
When the inversion center is shifted to the middle of two adjacent (001) layers, 
the inversion operator acquires a factor $e^{-iq_3}$, and the inversion eigenvalues 
switch signs at the TRI momenta with $q_{3} = \pi$. 
In this choice of inversion center, in addition to the band inversion at $(\pi,\pi,0)$, 
there are three band inversions at $(0,0,\pi)$, $(0,\pi,\pi)$, and $(\pi,0,\pi)$, 
as sketched in Fig.~\ref{fig7}(c). This implies the symmetry indicators $(001;0)$. 
In the reduced BZ as sketched in Fig.~\ref{fig7}(d), there is one band inversion 
at each TRI momentum in the $q_{3} = 0$ plane, 
and the symmetry indicators become $(000;0)$, which also characterizes $\alpha$-Bi$_4$I$_4$. 
 
Clearly, while the ${\mathbb Z}_2$ indices are not robust against the unit cell doubling, 
they remain the same under the inversion center shifting. 
By contrast, the ${\mathbb Z}_4$ index behaves in the opposite manner. 
Both Figs.~\ref{fig7}(a) and~\ref{fig7}(c) illustrate the band inversions of the $\beta$ phase WTI, 
since its inversion center can be placed either in a (001) layer or between two adjacent layers.
Moreover, Figs.~\ref{fig7}(b) and~\ref{fig7}(d) illustrate the band inversions 
of $\alpha$-Bi$_4$Br$_4$ and $\alpha$-Bi$_4$I$_4$, respectively, 
since the additional inter-layer couplings induced by the structure transitions are sufficiently weak 
compared with the bulk band gap of the $\beta$ phase WTI.
In fact, those weak couplings play two roles in determining 
the symmetry and topology of the two $\alpha$ phases.
First, they reduce the translational symmetry, opening the side surface band gaps.
Second, they reduce the inversion symmetry, fixing the inversion center locations. 
These two effects lead to ${\mathbb Z}_4=2$ for $\alpha$-Bi$_4$Br$_4$ 
and ${\mathbb Z}_4=0$ for $\alpha$-Bi$_4$I$_4$.
More fundamentally, they together give rise to the surface SSH models in Sec.~\ref{sec:SSH}
and the surface topological invariants in Sec.~\ref{invariant}.

\begin{figure}[t!]
\includegraphics[scale=1]{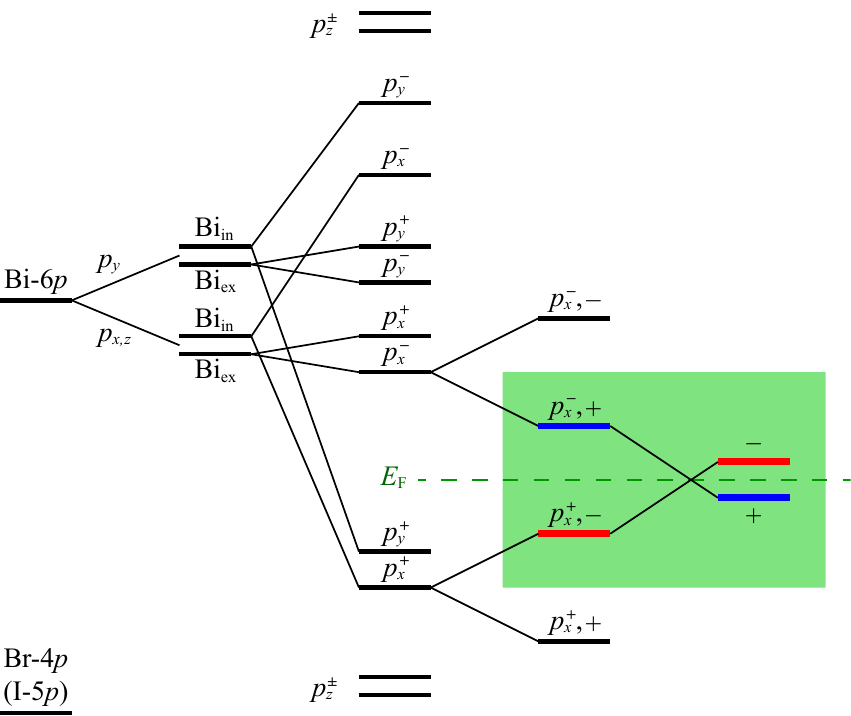}
\caption{The band evolution at the $L$ and $M$ points near the Fermi energies of Bi$_{4}$X$_{4}$. 
From left to right, we consider the electronegativity of the atomic orbitals, 
the splitting due to the (010) mirror symmetry, the formation of bonding and anti-bonding states 
as a result of the intra- and inter-chain couplings, 
and the possible band inversion driven by the SOCs. The details are explained in the text.} 
\label{fig8}
\end{figure}

\subsection{Band evolution of Bi$_{4}$X$_{4}$}\label{levels} 

To better understand the physical mechanism of the band inversions revealed in Sec.~\ref{sec:band}, 
now we explain the band evolution of Bi$_{4}$X$_{4}$ at various stages~\cite{Liu2016, Zhou2014}. 
We first elucidate the case for the $\beta$ phase and then generalize it to the $\alpha$ phase, 
as illustrated in Fig.~\ref{fig8}.

Due to the large electronegativity of halogen atoms, 
the bands near the Fermi energy are mainly determined by the $6p$ orbitals of bismuth atoms. 
First, consider the (010) mirror symmetry. 
The $p_{y}$ orbitals are decoupled from the $p_{x,z}$ orbitals 
since their mirror eigenvalues are different. Given the atomic chain orientation, 
the energies of $p_{y}$ orbitals are higher than those of $p_{x,z}$ orbitals; 
the hopping between $p_{y}$ orbitals is of $\sigma$-type, 
whereas that between $p_{x,z}$ orbitals is of $\pi$-type. 
Then, consider the inversion symmetry in a chain. 
${\rm Bi_{in/ex}}$ and ${\rm Bi'_{in/ex}}$ atoms are reflected to each other under inversion.
They form the bonding and anti-bonding states split in energy:  
$|{\rm Bi_{in/ex}}, p_{i}^{\pm}\rangle=(|{\rm Bi_{in/ex}}, p_{i}\rangle \pm |{\rm Bi'_{in/ex}}, p_{i}\rangle)/\sqrt{2}$,  
where $i=x,y,z$. 

Next, we count the stronger intra-chain couplings. 
Because of the short ${\rm Bi_{in}}$-${\rm Bi'_{in}}$ distance, 
the energy splittings between states $|{\rm Bi_{in}}, p_{x}^{\pm}\rangle$ 
and between states $|{\rm Bi_{in}}, p_{y}^{\pm}\rangle$ are large. On the contrary, 
the large  ${\rm Bi_{ex}}$-${\rm Bi'_{ex}}$ distance results in negligible energy splittings. 
However, states $|{\rm Bi_{ex}}, p_{x,y}^{\pm}\rangle$ can be coupled to states 
$|{\rm Bi_{in}}, p_{x,y}^{\pm}\rangle$ via the $\pi$-bonding and acquire splittings 
in the opposite fashion of $|{\rm Bi_{in}}, p_{x,y}^{\pm}\rangle$. Due to the $\sigma$-bonding 
between states $|{\rm Bi_{in}}, p_{z}^{\pm}\rangle$ and $|{\rm Bi_{ex}}, p_{z}^{\pm}\rangle$, 
the $p_{z}$ orbitals split and shift far away from the Fermi energy. 

Moreover, we take into account the weaker inter-chain couplings. 
As the chains are closer in the $\bm a$ direction than in the $\bm c$ direction, 
the energy splitting mainly occurs within each (001) layer due to the couplings in the $\bm a$ direction.
Two adjacent chains in the same layer can be related by the inversion symmetry, 
and their states $|{\rm Bi_{in/ex}}, p_{x}^{\pm}\rangle$ form the bonding and anti-bonding states   
$|{\rm Bi_{in/ex}}, p_{x}^{\pm},\pm\rangle$, where the new $\pm$ signs denote the inversion eigenvalues. 
As a results, the states $|{\rm Bi_{in}}, p_{x}^{+},-\rangle$ and $|{\rm Bi_{ex}}, p_{x}^{-},+\rangle$ 
become the valence and conduction bands closest to the Fermi energy, respectively. 

Finally, we include the effect of spin-orbit couplings (SOC). The SOCs mix the $p_{x}$ orbitals 
with the $p_{y,z}$ orbitals that have the same inversion eigenvalues. As a result, 
band inversions occur near the Fermi energy at either one or both of the two TRI momenta 
$L$ and $M$, and Bi$_{4}$X$_{4}$ become topologically nontrivial.

The above picture for the $\beta$ phase equally applies to $\alpha$-Bi$_{4}$Br$_{4}$, 
as its inversion center is also in a (001) layer. 
For $\alpha$-Bi$_{4}$I$_{4}$, as its inversion center is in the middle of two adjacent layers instead, 
the two pairs of $\pm$ signs above should be both interpreted as the labels of bonding and anti-bonding states, 
and the NN inter-chain couplings in the $\bm c$ direction needs to be further considered 
for states $|{\rm Bi_{in}}, p_{x}^{+},-\rangle$ and $|{\rm Bi_{ex}}, p_{x}^{-},+\rangle$ 
to form the inversion eigenstates that are eventually band-inverted by the SOCs. 

\section{Effective Tight-Binding Models}\label{models}

\begin{figure*}[ht!]
\includegraphics[scale=1]{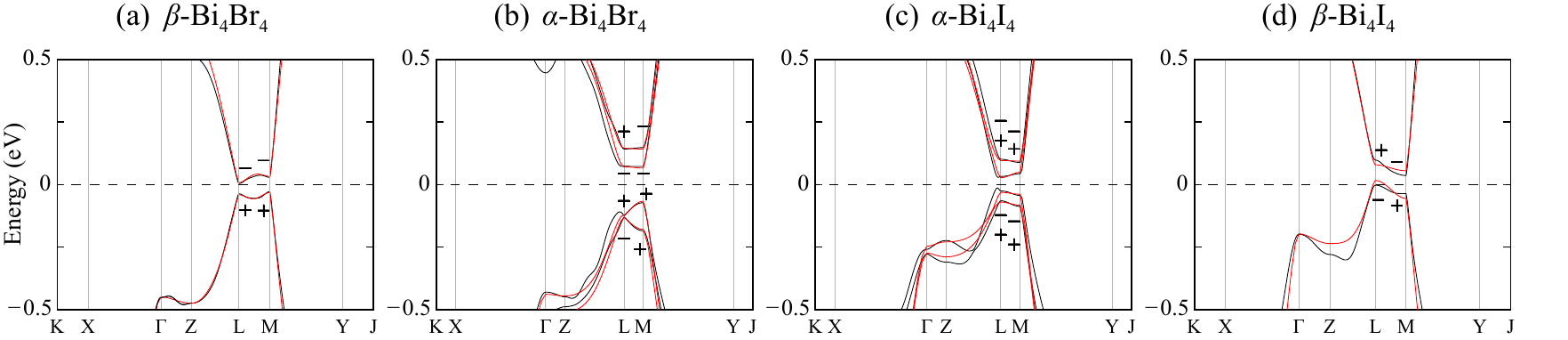}
\caption{The band structures (black) and band inversions ($\pm$) of (a) $\beta$-Bi$_4$Br$_4$, 
(b) $\alpha$-Bi$_4$Br$_4$, (c) $\alpha$-Bi$_4$I$_4$, and (d) $\beta$-Bi$_4$I$_4$ 
in Fig.~\ref{fig6} fitted by our effective tight-binding models (red) 
evaluated by the set of parameter values in Appendix~\ref{fitting}.} 
\label{fig9}
\end{figure*}

We construct the effective tight-binding model for both $\alpha$- and $\beta$-Bi$_{4}$X$_{4}$ 
based on their crystal and band structures revealed in Secs.~\ref{crystal} and~\ref{sec:band}. 
We start from the construction of the WTI model for the $\beta$ phase and then derive 
the models for the two distinct $\alpha$ phases by applying zone folding and Peierls distortion. 
With reasonable sets of parameter values in Appendix~\ref{fitting}, 
our models well fit the band inversions and band structures in Fig.~\ref{fig6}, 
as shown in Fig.~\ref{fig9}.
Imposing a topological boundary condition~\cite{Zhang2012}, 
we derive the surface states that resemble the SSH model 
and propose a surface topological invariant that explains 
the hinge state patterns in Figs.~\ref{fig1}-\ref{fig3}.  

\subsection{$\beta$-Bi$_{4}$X$_{4}$ model}

We choose a basis in which the three spatial symmetries discussed in Sec.~\ref{crystal}  
and the $\mathcal{T}$ symmetry can be specified as follows:
\begin{equation}\label{symmetry}
\mathcal{P}=\sigma_{z}, \;\; \mathcal{M}_{b} = is_{y}, \;\;
\mathcal{C}_{2} =\mathcal{M}_{b}\mathcal{P}, \;\; \mathcal{T}=i\mathcal{K}s_{y},
\end{equation}
where $\mathcal{K}$ is the complex conjugation. As such, 
the low-energy model for $\beta$-Bi$_{4}$X$_{4}$ can be built by using the four basis states 
$\ket{+,\uparrow}$, $\ket{+,\downarrow}$, $\ket{-,\uparrow}$, and $\ket{-,\downarrow}$, 
where $\pm$ and $\uparrow/\downarrow$ denote the eigenvalues of $\sigma_{z}$ and $s_z$, 
respectively. As illustrated in Fig.~\ref{fig8}, 
the $\ket{\sigma_z=\pm}$ states are mainly from the $p_{x}$ orbitals of ${\rm Bi_{in/out}}$ atoms, 
and the $\ket{s_z=\;\uparrow/\downarrow}$ states are dominated by the electron spins. 

We further assume that the four basis states are localized in (001) layers and consider only 
the NN (intra-layer) intra-chain, NN (intra-layer) inter-chain, and NN inter-layer (inter-chain) 
hopping processes, as well as the onsite potentials.
It follows that the symmetries in Eq.~(\ref{symmetry}) dictate the $\beta$ phase Hamiltonian to be 
\begin{align}\label{H-beta}
\!H^{\beta}\!=\!H^{L}\! &+ 2(d_c + m_c\sigma_{z}) \cos{q_3} 
+ 2(t_{c}\sigma_{x}s_{y} + t_{c}''\sigma_{y} ) \sin{q_3},\nonumber\\
H^{L}\!=\,&M\sigma_z + (t_{a} \sigma_{y} + t_{a}'' \sigma_{x}s_{y} ) (\sin {q_{1}} + \sin {q_{2}} )\nonumber \\
&+ D + ( t_{b} \sigma_{x}s_{z} + t_{b}'' \sigma_{x}s_{x}) \sin {(q_{2}-q_{1})},\nonumber\\
M=\,&\,m_0 + m_a (\cos {q_{1}} + \cos {q_{2}}) + m_b\cos{(q_{2}-q_{1})},\nonumber \\
D=\,&\,d_0 + d_a (\cos {q_{1}} + \cos {q_{2}}) + d_b\cos{(q_{2}-q_{1})},
\end{align}
where $H^{L}$ is the (001) monolayer Hamiltonian, $q_{i} = {\bm k} \cdot {\bm a_{i}}$,
and $\bm a_{1,2}=({\bm a} \mp {\bm b})/2$ and ${\bm a_{3}} = {\bm c}$ 
are the primitive lattice vectors shown in Fig.~\ref{fig5}. 

For simplicity, we set $t_{a}'' = t_{b}'' = t_{c}'' = 0$ in Eq.~(\ref{H-beta}) hereafter. 
We may interpret $t_{i} \gg t_{i}''$ ($i=a, b, c$) as follows. 
As the $\ket{\sigma_z=\pm}$ states are mainly from the $p_{x}$ orbitals, 
spin independent hopping processes dominate in the $\bm a$ direction, i.e., $t_{a} \gg t_{a}''$.
As the contribution of $p_{z}$ orbitals is negligibly weak at the Fermi energy, 
the SOC terms $\propto s_{z}$ dominate those $\propto s_{x,y}$, i.e., $t_{b} \gg t_{b}''$. 
Near the $L$ and $M$ points in Fig.~\ref{fig6}(a), 
the bands in the absence of the SOCs become much steeper when the SOC effect is taken into account,  
i.e., $t_{c} \gg t_{c}''$. 
 
To derive the $\alpha$ phase models, we apply zone folding to Eq.~(\ref{H-beta}) 
by doubling the unit cell in the $\bm c$ direction. 
This model with two (001) layers per unit cell reads
\begin{align}\label{H-beta1}
\bar{H}^{\beta}=\;&H^{L}  
+ 2(d_c + m_c\sigma_{z})\tau_x\cos \frac{q_3}{2} \nonumber \\
&+ 2t_{c}\sigma_{x}s_{y}\tau_x\sin \frac{q_3}{2},
\end{align}
where $\tau_z=\pm$ denote the even and odd (001) layers.
Note that in this model the inversion operator becomes 
$\mathcal{P}=\sigma_z\tau_x$ for inversion center placed in the middle of two adjacent layers
and remains $\mathcal{P}=\sigma_z$ for inversion center placed in a layer.
In order to make this model periodic in $q_3$, we further perform the gauge transformation 
in Appendix~\ref{GT} and obtain 
\begin{align}\label{H-beta2}
\tilde{H}^{\beta}=\;&H^{L} 
+ (d_c + m_c\sigma_{z}) [\tau_x + (\tau_x\cos{q_3}+\tau_y\sin{q_3})] \nonumber \\
&+ t_{c}\sigma_{x}s_{y} [\tau_y-(\tau_y\cos{q_3}-\tau_x\sin{q_3})],
\end{align}
in which the inversion operator remains $\mathcal{P}=\sigma_z\tau_x$ 
for inversion center placed in the middle of two adjacent layers and becomes 
$\mathcal{P}=\sigma_z(1+\tau_z)/2+e^{iq_3}\sigma_z(1-\tau_z)/2$ for inversion center placed in a layer. 
Note that ${\bm a_{3}} = {2\bm c}$ in both Eqs.~(\ref{H-beta1})~and~(\ref{H-beta2}). 

For the models in Eqs.~(\ref{H-beta})-(\ref{H-beta2}) to describe the WTI with band inversions 
at the $M$ and $L$ points, $(\pi,\pi,0)$ and $(\pi,\pi,\pi)$, $M_{\pi,\pi}<-2|m_c|$ and 
$M_{0,0},\,M_{\pi,0}>2|m_c|$ are dictated based on the Fu-Kane criterion~\cite{Fu2007}. 
The models for $\alpha$-Bi$_{4}$Br$_{4}$ and $\alpha$-Bi$_{4}$I$_{4}$ are derived below 
by introducing additional symmetry-allowed terms to the zone-folded $\beta$ phase model. 
From the bulk perspective, these terms may seem secondary, as they simply shift the energies of bands 
and remove the accidental degeneracies of the zone-folded $\beta$ phase model;  
when their energy scale is smaller than the gap along the $ML$ line  
$\sim|M_{\pi,\pi}\pm 2m_c|$, the two $\alpha$ phase models are equivalent to 
the zone-folded $\beta$ phase model, as shown in Sec.~\ref{doubling}. 
More substantially, these terms reduce the translational and inversion symmetries by choosing a location for the inversion center.
From the surface perspective, these terms are essential. They open the side surface band gaps in Sec.~\ref{dimerization}, 
yield the surface SSH models in Sec.~\ref{sec:SSH}, and validate the surface topological invariants in Sec.~\ref{invariant}. 

\subsection{$\alpha$-Bi$_{4}$Br$_{4}$ model}

In contrast to the $\beta$ phase, the primitive unit cell of $\alpha$-Bi$_{4}$Br$_{4}$ 
consists of two (001) layers, and its inversion center can only be placed in a layer.
Thus, Eq.~(\ref{H-beta1}) provides a reasonable, convenient starting point to construct 
the effective tight-binding model for $\alpha$-Bi$_{4}$Br$_{4}$. 
Moreover, the even and odd (001) layers are nearly (001) mirror images of each other. 
Given $s_z H^{L} s_z = H^{L}$, the crystal symmetries, and the orbital characters, 
with the zeroth order corrections,  
$\alpha$-Bi$_{4}$Br$_{4}$ can be described by 
\begin{align}\label{H-alpha1}
\bar{H}^{\alpha}_{\rm BiBr}=\;&\bar{H}^{\beta} + (d_0' + m_0'\sigma_{z})\tau_z
+ 2t_c'\sigma_{y}s_{y}\tau_y\sin \frac{q_3}{2}\nonumber\\ 
&+ 2(d_c' + m_c'\sigma_{z}) s_y\tau_y\cos\frac{q_3}{2},
\end{align}
for which $\mathcal{P}=\sigma_z$. In Eq.~(\ref{H-alpha1}) 
the terms $\propto\tau_z$ characterize the differences between the even and odd layers, 
and the terms $\propto\tau_y$ are new SOCs.
To make this model periodic in $q_3$, 
we further perform the gauge transformation in Appendix~\ref{GT} and obtain 
\begin{align}\label{H-alpha3}
\!\!\!{H}^{\alpha}_{\rm BiBr}\!=\,&\tilde{H}^{\beta} + (d_0' + m_0'\sigma_{z})\tau_z\nonumber\\
&+ t_c'\sigma_{y}s_y [-\tau_x + (\tau_x\cos{q_3}+\tau_y\sin{q_3})] \nonumber \\
&+ (d_c' + m_c'\sigma_{z})s_y[\tau_y+(\tau_y\cos{q_3}-\tau_x\sin{q_3})],\!
\end{align}
for which $\mathcal{P}=\sigma_z(1+\tau_z)/2+e^{iq_3}\sigma_z(1-\tau_z)/2$.

\subsection{$\alpha$-Bi$_{4}$I$_{4}$ model}

Although the primitive unit cell of $\alpha$-Bi$_{4}$I$_{4}$ also consists of two (001) layers, 
different from $\alpha$-Bi$_{4}$Br$_{4}$, 
its inversion center can only be placed in the middle of two adjacent layers. 
For this reason, while the even and odd layers are related to each other, 
the inter-layer spacings become not uniform any more.  
In this case, the inter-layer couplings become alternating in $\alpha$-Bi$_{4}$I$_{4}$, 
and Eq.~(\ref{H-beta2}) is a more convenient starting point to construct its effective tight-binding model.
Considering the alternating couplings, we find that $\alpha$-Bi$_{4}$I$_{4}$ can be described by 
\begin{align}\label{H-alpha2}
{H}^{\alpha}_{\rm BiI}=\;&H^{L} + t \sigma_x\tau_z + t' \sigma_y s_y\tau_z 
+ [(d_c + m_c\sigma_{z})\tau_x\nonumber\\
&+ (d_c' + m_c'\sigma_{z}) (\tau_x\cos{q_3}+\tau_y\sin{q_3})]\nonumber \\
&+ \sigma_{x}s_{y} [t_c \tau_y - t_c' (\tau_y \cos{q_3}-\tau_x \sin{q_3})],
\end{align}
for which $\mathcal{P}=\sigma_z\tau_x$. In Eq.~(\ref{H-alpha2}), the terms $\propto\tau_z$
are extra symmetry-allowed zeroth order corrections and characterize the differences between 
the even and odd layers. 
Note that Eq.~(\ref{H-alpha2}) does not count the relative shifts between adjacent layers 
within the $\bm a$-$\bm b$ plane. Appendix~\ref{accurate} provides 
a more accurate model that takes into account this ignored effect. 
In fact, the more accurate model reduces to Eq.~(\ref{H-alpha2}) near the $ML$ line. 

\subsection{Surface SSH dimerization}\label{dimerization}

Following the ZKM theory~\cite{Zhang2012,Zhang2013}, we can impose the topological boundary conditions 
in the $(100)$ and $(\bar100)$ directions and obtain the surface states for the two cases. 
This allows us to derive the surface SSH models in Fig.~\ref{fig3}, 
formulate a surface topological invariant determining the hinge state patterns in Figs.~\ref{fig1} and~\ref{fig2}, 
and characterize the topological distinctions between the two $\alpha$ phases.

We start from the $\beta$ phase WTI that has two band inversions, 
one at $M$ point $(\pi,\pi,0)$ and one at $L$ point $(\pi,\pi,\pi)$. Near the $ML$ line, 
Eq.~(\ref{H-beta}) yields the following $(100)$ and $({\bar 1}00)$ surface Hamiltonians 
\begin{align}\label{H-s-beta}
h^{\beta} = D_{\pi,\pi} + \eta t_{b}s_{z}q_b + 2d_c\cos{q_3} + 2\eta t_{c}s_{y}\sin{q_3},
\end{align}
where $D_{\pi,\pi}=d_0-2 d_a + d_b$.
Note that the $(100)$ and $({\bar 1}00)$ surface states 
are the eigenstates of $\sigma_x$ with $\eta=\pm$ eigenvalues, respectively~\cite{Zhang2012}. 
As a result, those terms anticommuting with $\sigma_x$ in Eq.~(\ref{H-beta}) only produce
hybridization between the two surfaces~\cite{Zhang2013}, 
which is negligibly weak when the two surfaces are well separated,  
and can thus be safely ignored in Eq.~(\ref{H-s-beta}). 
In fact, the two surface models in Eq.~(\ref{H-s-beta}) can be constructed by directly considering 
the symmetries in Eq.~(\ref{symmetry}) and the orbitals in Fig.~\ref{fig8}.
Each surface model in Eq.~(\ref{H-s-beta}) describes two connected gapless Dirac cones 
respecting $\mathcal{M}_b$ symmetry, 
one at $(0,0)$ and the other at $(0,\pi)$ with different Dirac-point energies.  
The two surface models in Eq.~(\ref{H-s-beta}) can be related by $\mathcal{P}$ or $\mathcal{C}_{2}$ symmetry. 

Similar to Eq.~(\ref{H-beta2}), we apply zone folding to Eq.~(\ref{H-s-beta}) 
by doubling the unit cell in the $\bm c$ direction and obtain 
\begin{align}\label{H-s-beta2}
\!\!{\tilde h}^{\beta} =\; & D_{\pi,\pi} + \eta t_{b}s_{z}q_b 
+ d_c [\tau_x + (\tau_x\cos{q_3}+\tau_y\sin{q_3})] \nonumber \\
&+ \eta t_{c}s_{y} [\tau_y-(\tau_y\cos{q_3}-\tau_x\sin{q_3})].
\end{align}
It is clear that the two Dirac surface states in Eq.~(\ref{H-s-beta}) are folded into 
$(0,0)$ in Eq.~(\ref{H-s-beta2}), and that they remain gapless. 
Moreover, from the terms $\propto\tau_{x,y}$ in Eq.~(\ref{H-s-beta2}), 
the inter-layer tunneling-up and -down matrices read 
\begin{align}\label{T-beta}
T^{\beta,\pm}=d_c\mp i \eta s_y t_c,
\end{align}
which implies no surface dimerization since $|T^{\beta,+}|=|T^{\beta,-}|$. 
The gapless nature and the absence of dimerization agree well with  
the edge construction in Sec.~\ref{sec:SSH} that the WTI side surface models are 
analogous to the critical point of the SSH model.
 
In the same fashion, Eqs.~(\ref{H-alpha3}) and~(\ref{H-alpha2}) respectively yield 
the $(100)$ and $({\bar 1}00)$ surface Hamiltonians of $\alpha$-Bi$_{4}$Br$_{4}$
\begin{align}\label{H-s-alpha1}
\!\!\!\!{h}^{\alpha}_{\rm BiBr} = {\tilde h}^{\beta} 
+ d_c' s_y [\tau_y+(\tau_y\cos{q_3}-\tau_x\sin{q_3})] + d_0'\tau_z,\!\!\!
\end{align}
with the inter-layer tunneling-up and -down matrices
\begin{align}\label{T-alpha1}
T_{\rm BiBr}^{\alpha,\pm}=d_c - i s_y (d_c'\pm \eta t_c) ,
\end{align}
and the $(201)$ and $({\bar 2}0{\bar 1})$ surface Hamiltonians of $\alpha$-Bi$_{4}$I$_{4}$
\begin{align}\label{H-s-alpha2}
\!\!\!\!{h}^{\alpha}_{\rm BiI} = \;& D_{\pi,\pi} + \eta t_{b}s_{z}q_b + [d_c\tau_x + d_c'(\tau_x\cos{q_3}+\tau_y\sin{q_3})] \nonumber \\
&+ \eta s_{y} [t_c\tau_y-t_c'(\tau_y\cos{q_3}-\tau_x\sin{q_3})] + \eta t \tau_z,
\end{align}
with the inter-layer tunneling-up and -down matrices
\begin{align}\label{T-alpha2}
T_{\rm BiI}^{\alpha,+}=d_c - i \eta s_y t_c ,\quad T_{\rm BiI}^{\alpha,-}=d_c' + i \eta s_y t_c' .
\end{align}

Evidently, surface dimerization is present in Eqs.~(\ref{H-s-alpha1}) and~(\ref{H-s-alpha2}),
as $|T_{\rm BiBr}^{\alpha,+}|\neq|T_{\rm BiBr}^{\alpha,-}|$ and 
$|T_{\rm BiI}^{\alpha,+}|\neq|T_{\rm BiI}^{\alpha,-}|$. 
Similarly in the two cases, the dimerization gaps the two surface states at $(0,0)$ in Eq.~(\ref{H-s-beta2}). 
Differently, the dimerization is characterized by $\eta t_c d_c'$ in Eq.~(\ref{H-s-alpha1}) 
and by $|d_c|-|d_c'|$ and $|t_c|-|t_c'|$ in Eq.~(\ref{H-s-alpha2});
the former only exists at the surfaces, whereas the latter even exists in the bulk.
Moreover, for the two spins ($s_y=\pm$) the dimerization patterns are the same in both cases, 
however, for the two side surfaces ($\eta=\pm$) 
the dimerization patterns are the same for $\alpha$-Bi$_{4}$I$_{4}$ 
but opposite for $\alpha$-Bi$_{4}$Br$_{4}$. 
All these results agree well with the edge construction in Sec.~\ref{sec:SSH} 
and the atomic dimerization in Sec.~\ref{ADim}.
  
\subsection{Surface topological invariants}\label{invariant}

From the analysis above, it is clear that a mirror (or $s_y$-resolved) 
winding number~\cite{Zhang2013a} can be used to characterize the topological properties 
of the two surface models in Eqs.~(\ref{H-s-alpha1}) and~(\ref{H-s-alpha2}). 
This is allowed because in the $\bm c$ direction these surface models are 1D tight-binding models. 
Applying the results in Appendix~\ref{winding} here and considering the terms $\propto \tau_{x,y}$ at $q_b=0$ only, 
we obtain
\begin{align}\label{wn1}
\gamma_{\rm BiBr}^{\eta,s_y} = \Theta\left(-\eta t_c d_c'\right) 
\end{align}
for the model in Eq.~(\ref{H-s-alpha1}) and 
\begin{align}\label{wn2}
\gamma_{\rm BiI}^{\eta,s_y} = \Theta\left(\sqrt{{d_c'}^2+{t_c'}^2}-\sqrt{{d_c}^2+{t_c}^2}\right)
\end{align}
for the model in Eq.~(\ref{H-s-alpha2}), 
with $\Theta$ the Heaviside function.

For both $\alpha$-Bi$_{4}$Br$_{4}$ and $\alpha$-Bi$_{4}$I$_{4}$, 
their winding numbers are independent of $s_y$.  
This implies that the topological invariants and dimerization patterns are robust against mirror symmetry breaking.
At the two opposite side surfaces, 
the winding numbers are different for $\alpha$-Bi$_{4}$Br$_{4}$ 
but the same for $\alpha$-Bi$_{4}$I$_{4}$. 
Both conclusions are consistent with the dimerization analyses in Secs.~\ref{sec:SSH},~\ref{ADim}, and~\ref{dimerization}. 
Although a winding number is gauge dependent, 
the differences in winding number between the two opposite side surfaces (of the same $\alpha$ phase) 
and between the two different $\alpha$ phases are both gauge invariant.  

The terms $\propto \tau_z$ in Eqs.~(\ref{H-s-alpha1}) and (\ref{H-s-alpha2}) break the chiral symmetry and 
seem to make the winding numbers in Eqs.~(\ref{wn1}) and (\ref{wn2}) meaningless. 
In general, as the chiral symmetry breaking produces particle-hole asymmetry,   
a zero-energy bound state implied by a nontrivial winding number is not necessarily 
pinned to the middle of band gap and can be removed perturbatively. 
However, this is not the case here. 
The pairs of zero modes implied by the nontrivial winding numbers 
become dispersive across the surface band gaps in the presence of 
the terms $\propto q_b$~\footnote{This is inherited from the monolayer TIs. 
In other words, terms $\propto\sin q_b$ can not ensure the same protection.} 
and form helical hinge states with gapless Dirac points at $q_b=0$, which is protected by the $\mathcal{T}$ symmetry.
 
\subsection{Domain wall argument} 

Although the dimerization analyses and the topological invariants above are sufficient 
to understand the helical hinge states of the two $\alpha$ phases, 
here we briefly discuss how to use the ZKM theory~\cite{Zhang2012,Zhang2013} 
to construct domain walls (DW) and deduce the presence or absence of hinge states. 
This DW method has been employed to obtain the original higher-order chiral TI~\cite{Zhang2013} 
and to demonstrate several more recent examples~\cite{Zhang2019a, Schindler2018a, Song2017, Schindler2018, Langbehn2017, Fang2019, Yan2018, Wang2018, Khalaf2018, Sheng2019}. 
 
\begin{figure}[t!]
\includegraphics[scale=1]{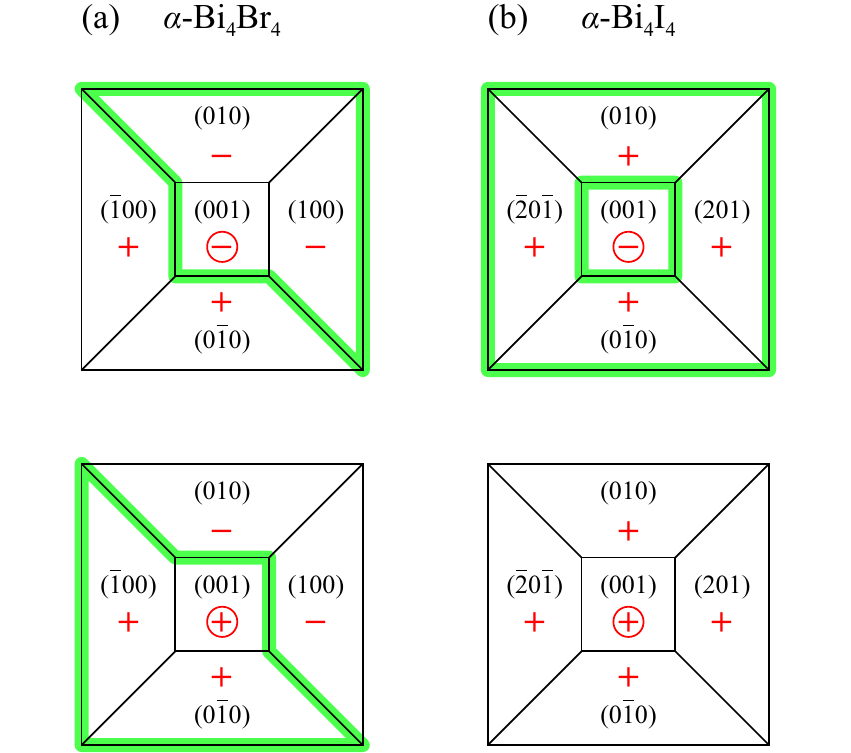}
\caption{Schematics of inversion symmetric (a) $\alpha$-Bi$_{4}$Br$_{4}$ and 
(b) $\alpha$-Bi$_{4}$I$_{4}$ samples with or without helical hinge states. 
The relative signs of the mass term gapping the two surface Dirac cones are indicated in red. 
The hinge states along the domain walls switching the mass signs are indicated in green. 
(The other two hinge state patterns for $\alpha$-Bi$_{4}$Br$_{4}$ are not shown. 
The $\mathcal{C}_2$ symmetry is not considered here.)} 
\label{DW}
\end{figure}

Consider the local frame of a generic surface formed by a vector $\bm{j}$ along the hinge, 
the outward surface normal $\bm k$, and the direction $\bm i=\bm j \times \bm k$ (at the surface). 
While the $\bm j$ axis is fixed, the $\bm k$ and $\bm i$ axes co-move with the surface.
As the bulk translational symmetry is broken in the $\bm k$ direction by a topological boundary condition, 
we can derive a surface Hamiltonian that describes all the surfaces 
sharing the same hinge in the $\bm j$ direction~\cite{Zhang2012}. 
For a generic surface, the two resulting surface Dirac cones are gapped due to a hybridization mass. 

For $\alpha$-Bi$_{4}$Br$_{4}$, viewed in the local frame, 
the $\mathcal{P}$ symmetry dictates the mass term to have the opposite signs at two opposite surfaces. 
This is because the dimerization of $\alpha$-Bi$_{4}$Br$_{4}$ is a surface effect 
as revealed in Sec.~\ref{dimerization}. As a result, 
depending on the relative sign of the mass term at the top surface, 
an inversion symmetric sample exhibits one loop of hinge states in one of the two fashions, 
as depicted in Fig.~\ref{DW}(a). 
(The mass term at the $(010)$ or $(0\bar{1}0)$ surface is odd under the $\mathcal{C}_2$ rotation, 
restoring the two gapless Dirac cones.)
For $\alpha$-Bi$_{4}$I$_{4}$, however, 
the $\mathcal{P}$ symmetry dictates the mass term to have the same sign at two opposite surfaces. 
In fact, the mass term has the same sign for all the side surfaces, 
since the dimerization of $\alpha$-Bi$_{4}$I$_{4}$ is a bulk effect as revealed in Sec.~\ref{dimerization}.
It follows that, depending on the relative sign of the mass term at the top surface, 
an inversion symmetric sample hosts either no hinge states or two loops of hinge states, 
as depicted in Fig.~\ref{DW}(b). These are consistent with the results presented 
in Figs.~\ref{fig1}(a),~\ref{fig1}(c),~\ref{fig2}(a), and~\ref{fig2}(c).

\section {Discussion and Experiment}\label{discussion} 

Although both are HOTIs with helical hinge states, 
$\alpha$-Bi$_{4}$Br$_{4}$ is more intrinsic (bulk-obstructed) 
whereas $\alpha$-Bi$_{4}$I$_{4}$ is more extrinsic (boundary-obstructed)~\cite{Geier2018, Khalaf2019}.  
As featured in Figs.~\ref{fig1} and~\ref{fig2}, attaching a 2D TI to the (001) surface 
can annihilate an existing hinge state of $\alpha$-Bi$_{4}$I$_{4}$ but not that of $\alpha$-Bi$_{4}$Br$_{4}$.
Markedly, their clear distinction in the hinge state pattern highlights 
the critical role played by the location of inversion center that has so far been overlooked in the literature.    
Fundamentally, as only the $\mathcal{T}$ symmetry is required to protect a local 1D helical mode, 
the helical hinge states are robust against the $\mathcal{P}$ symmetry breaking, 
as long as the perturbation neither closes the gaps of different surfaces nor hybridizes the states at different hinges. 
This is in sharp contrast to the case of topological mirror insulators and superconductors~\cite{Hsieh2012, Zhang2013a}, 
in which the mirror symmetry breaking exclusively removes the gapless boundary modes.
Given that the symmetry indicators of $\alpha$-Bi$_{4}$I$_{4}$ are completely trivial~\cite{Zhang2019, Vergniory2019, Tang2019},
our results imply that there are likely to be many topological materials 
beyond the scope of symmetry indicators and awaiting to be discovered. 

We stress that the effective models for the side surface states in Sec.~\ref{models} 
are continuum models in the chain direction but lattice models in the stacking direction. 
As the side surface states can be constructed by the helical edge states of (001) monolayers that are 2D $\mathbb{Z}_2$ TIs, 
it does not exist any lattice model to account for the surface states in the chain direction. 
However, this is not the case in the stacking direction, 
as there are nevertheless dual surface Dirac cones that are gapless in the WTI case but gapped in the HOTI case. 
Best displayed by the  quasi-1D Bi$_{4}$X$_{4}$, 
this special feature reflects the intimate relations between the HOTI, WTI, and 2D TIs. 

Given that the (001) monolayers of the four materials are all 2D $\mathbb{Z}_2$ TIs as shown in Fig.~\ref{fig6},
one may wonder whether their (100) ((201) for $\alpha$-Bi$_{4}$I$_{4}$) films are also TIs.
Our MLWF-based calculations and effective tight-binding models both indicate that 
their monolayers and the $\beta$-Bi$_{4}$X$_{4}$ film of any thickness are $\mathbb{Z}_2$ trivial.  
However, any $\alpha$-Bi$_{4}$Br$_{4}$ (100) film thicker than three layers~\footnote{
The thinest (100) TI film is the tetralayer in our MLWF-based computation.  
It is the hexalayer instead in our effective tight-binding model calculation. 
The latter can be adjusted to match the former 
without affecting the bulk and surface band topologies by changing 
$t_a$ from 181.67~meV to 140.00~meV in Table~\ref{Table:parameters}.}
is a $\mathbb{Z}_2$ TI.
This can be understood by the top panels of Fig.~\ref{fig1}: 
each scenario has one hinge state at the top surface and one at the bottom. 
As the (100) thickness decreases, the two hinge states turn into the two edge states, 
as long as the inter-edge coupling in a (001) layer is sufficiently small compared with the inter-edge coupling between two (001) layers. 
On the contrary, an $\alpha$-Bi$_{4}$I$_{4}$ (201) film of any thickness is $\mathbb{Z}_2$ trivial. 
As each surface in the top panels of Fig.~\ref{fig2} has either zero or two hinge states, 
for a (201) film either there is no edge state or the dual edge states acquire a hybridized gap, though, which is small for a thick film.
This offers a different perspective on the distinction between the two HOTIs.

A variety of experiments can be carried out to examine our predictions.
The structure transition between the high-temperature $\beta$ and low-temperature $\alpha$ phases 
can be characterized by the resistance discontinuity and its hysteresis in bulk transport~\cite{Note101}  
and by the gapless or gapped nature of side surface states in ARPES~\footnotetext[104]{Ming Yi (private communication).}\cite{Autes2016, Noguchi2019, Note104}, 
in addition to single crystal X-ray diffraction, scanning tunneling microscopy (STM)~\cite{Schindler2018a}, 
and transmission electron microscopy (TEM). 
For different phases, their unique Lifshitz transitions in the surface states can be revealed by ARPES. 
For the two $\alpha$ phases, the hidden dimerization can be unambiguously determined by STM and TEM, 
and the hinge states can be directly imaged by microwave impedance microscopy (MIM)~\cite{Lai2011, Ma2015, Wu2016, Shi2019}. 
For odd (001) layers, the quantum spin Hall effect can be detected in edge transport, 
and the edge states can be mapped by MIM. 
(Recently, the observations of $\alpha$-Bi$_{4}$Br$_{4}$ hinge/edge states in 
ARPES and infrared absorption spectroscopy have been reported~\footnotetext[105]{Yugui Yao (private communication).}\cite{Noguchi2020, Note105}. Based on our results here,  
what have been observed in these two experiments appear to be the gapped (100) surface states.)

A plethora of hinge state signatures can be obtained in a gate-tunable multi-terminal device.  
In such a device, the Fermi energy can be tuned by a gate voltage into the bulk, surface, and hinge states.  
(i) The charge neutrality point is set by the Dirac points of gapless hinge states. 
(ii) While the surface states have a much larger density of states (DOS), 
a 1D helical mode has a constant DOS, i.e., $1/\pi\hbar v_F$.
The total DOS in the surface state gap can be used to estimate the number of hinges, 
step edges, or/and stacking faults that host 1D helical modes. 
(iii) The sign of Hall coefficient can indicate whether the Fermi energy crosses 
the electron or hole surface band and infer the size of surface state gap. 
(iv)  In the surface state gap, nonlocal conductances can be analyzed to determine 
the hinge state pattern and layer stacking order.  
(v) From weak antilocalization in magnetotransport, 
the temperature dependence of dephasing lengths can be extracted. 
The Nyquist length $L_n$ and the phase coherence length $L_{\varphi}$ should both scale as  
$T^{-1/3}$, $T^{-1/2}$, and $T^{-3/4}$ for the hinge, surface, and bulk states, 
respectively~\footnotetext[103]{C. N. Lau (private communication).}~\cite{Altshuler1985, Gehring2015, Note103}. 
(vi) The spatial distribution of surface and hinge conductances can be mapped by the aforementioned MIM. 
(vii) The helical hinge states have Fermi velocities $\sim3.5-6\times 10^5$~m/s~\footnote{
For $\alpha$-Bi$_{4}$Br$_{4}$ hinge states,
the Fermi velocities are $4.19\times 10^5$~m/s in Fig.~\ref{fig1}(a) and $5.58\times 10^5$~m/s in Fig.~\ref{fig1}(c). 
For $\alpha$-Bi$_{4}$I$_{4}$ hinge states,
the Fermi velocities are $4.63\times 10^5$~m/s and $3.79\times 10^5$~m/s in Fig.~\ref{fig2}(b).} 
and confinement lengths $<1$~nm, 
similar to the case of bismuthene on SiC(0001)~\cite{Stuhler2020}. 
The anticipated helical Tomonaga-Luttinger liquid behavior can be probed via 
the power-law dependences on energy and temperature in tunneling spectroscopy~\cite{Deshpande2010, Li2015, Stuhler2020}.

Our results establish a new TI physics paradigm and a unique quasi-1D material platform 
for exploring the interplay of geometry, symmetry, topology, and interaction. 
Besides the discussed experiments,
studies on the coupling to ferromagnet, superconductor, or their linear junctions, 
the possible topological phase transitions under strain~\cite{Liu2016},  
and the influence of electron-electron interactions would be extremely interesting. 
 
\begin{acknowledgments}
F.Z. is grateful to Benjamin Wieder, Hongming Weng, Xiangang Wan, Marc Bockrath, Joseph Heremans, 
Bing Lv, Ming Yi, Chun Ning Lau, and Robert Birgeneau for insightful discussions.
C.Y., C.C.L., and F.Z. acknowledge the Texas Advanced Computing Center (TACC) for providing resources 
that have contributed to the research results reported in this work. 
This work was supported by National Science Foundation (NSF) under Grant No. DMR-1921581 through the DMREF program (F.Z.), 
Army Research Office (ARO) under Grant No. W911NF-18-1-0416 (F.Z.), 
UTD research enhancement fund (C.Y., C.C.L., and F.Z.), 
National Research Foundation of Korea under Grant Nos. 2016H1A2A1907780 
through the Global PhD Fellowship Program (C.Y.) and 2018R1A2B6007837 (C.Y. and H.M.), 
and SNU Creative Pioneering Researchers Program (C.Y. and H.M.).
\end{acknowledgments}

\appendix
\begin{table*}[t!] 
\caption{Crystal structure data of Bi$_{4}$X$_{4}$.
For $\alpha$-Bi$_{4}$I$_{4}$, the second line gives the values of the angle $\theta_{2}'$ 
and the distance Bi$_{\rm in}'$-Bi$_{\rm ex}'$. For $\alpha$-Bi$_{4}$Br$_{4}$, 
the two lines list the parameter values of the two symmetry unrelated (001) layers, respectively. 
Note that $\alpha=\gamma=90^{\circ}$ and that conventional unit cells are used.}
\label{table:crystal} 
\vspace{0.1cm}
\begin{tabular}{l|rrrr|cccc}
\hline\hline
\multicolumn{1}{c|}{
$\;\;\;\;$Phase$\;\;\;\;$} &
$\;\;$ $a$ ($\rm \AA$)$\;\!$ &
$\;\;\;$$b$ ($\rm \AA$) &
$\;\;\;\:\:$$c$ ($\rm \AA$)$\;$ &
$\;\;\;\:\!$ $\beta$ ($^{\circ}$)$\;\;\:$ &
$\;\;\,$$\theta_{1}$ ($^{\circ}$)$\,\;\;$ &
$\;$$\theta_{2}$ ($^{\circ}$)$\;$  &
$\;$ ${\rm Bi}_{\rm in}$-${\rm Bi}_{\rm in}'$ ($\rm \AA$)$\;$ &
$\;\;$${\rm Bi}_{\rm in}$-${\rm Bi}_{\rm ex}$ ($\rm \AA$)$\;\;$ \\
\hline
$\;\;$$\beta$-Bi$_{4}$Br$_{4}$   & 13.307 & 4.338 & 10.191   & 107.87$\;\;$     & 91.12   & 91.62   & 3.038   &  3.063       \\
\hline
$\;\;$$\alpha$-Bi$_{4}$Br$_{4}$  & 13.064 & 4.338 & 20.061   & 107.42$\;\;$    & 91.34   & 91.28   & 3.032   &  3.056        \\
                                                    &             &           &               &                          & 91.51   & 93.07  &  3.028  &   3.052       \\
\hline
$\;\;$$\alpha$-Bi$_{4}$I$_{4}$   & 14.245 & 4.428 & 19.968    & 87.04$\;\;$       & 93.49   & 92.80    & 3.040   &  3.057      \\
$\;\;\;\;\;$\cite{Note102}                                                   &             &           &                &                          &             & 92.39    &             &  3.044      \\
\hline
$\;\;$$\beta$-Bi$_{4}$I$_{4}$     & 14.386 & 4.430 & 10.493    & 107.87$\;\;$     & 93.18   & 92.42   & 3.049   & 3.071       \\
\hline\hline
\end{tabular}
\end{table*}

\section{Crystal structure data}\label{data}

Table~\ref{table:crystal} summarizes the crystal structure data of $\beta$-Bi$_{4}$Br$_{4}$, 
$\beta$-Bi$_{4}$I$_{4}$, $\alpha$-Bi$_{4}$Br$_{4}$, and $\alpha$-Bi$_{4}$I$_{4}$.
  
\section{Model fitting}\label{fitting}

Table~\ref{Table:parameters} summarizes a set of parameter values for the Bi$_{4}$X$_{4}$ models in Sec.~\ref{models} 
that can well fit the band inversions and band structures of the MLWF data, as shown in Fig.~\ref{fig9}. 
Below we describe our fitting procedure.

We first fit the intra-layer terms in $H^{L}$ of Eq.~(\ref{H-beta}) to the (001) monolayer Hamiltonians 
derived from the MLWF Hamiltonians by ignoring the inter-layer tunnelings. 
Informed by the bottom panels of Fig.~\ref{fig6}, $H^{L}$ has a band inversion at $(\pi,\pi)$.
Those terms even in momentum are fixed by the band energies at the TRI momenta, 
and those odd are determined by the band dispersions near $(\pi,\pi)$. 
For the $\alpha$ phases, those terms $\propto\tau_{z}$ are identified  
by the differences between the even and odd (001) layers, 
e.g., the direct band gaps of Bi$_{4}$Br$_{4}$ and the Dirac point energies of Bi$_{4}$I$_{4}$.
On top of these, we then fit the inter-layer tunnelings in Eqs.~(\ref{H-beta}),~(\ref{H-alpha3}), and~(\ref{H-alpha2}) 
with the band inversions and band energies at the $L$ and $M$ points. 

\section{Gauge transformation}\label{GT}

The transformation from the non-periodic Hamiltonians Eqs.~(\ref{H-beta1}) and~(\ref{H-alpha1}) 
to the periodic ones Eqs.~(\ref{H-beta2}) and~(\ref{H-alpha3}) may be specified as follows:
\begin{eqnarray}
\tilde{H}^{\beta} = U\bar{H}^{\beta}U^{-1},\quad {H}^{\alpha}_{\rm BiBr} = U\bar{H}^{\alpha}_{\rm BiBr}U^{-1},
\end{eqnarray}
where $U=(1+\tau_z)/2+e^{iq_3/2}(1-\tau_z)/2$, and $\tau_{z} = \pm 1$ denote the two layers in a unit cell. 

\section{More accurate $\alpha$-Bi$_{4}$I$_{4}$ model}\label{accurate}

\begin{figure}[b!]
\includegraphics[scale=1]{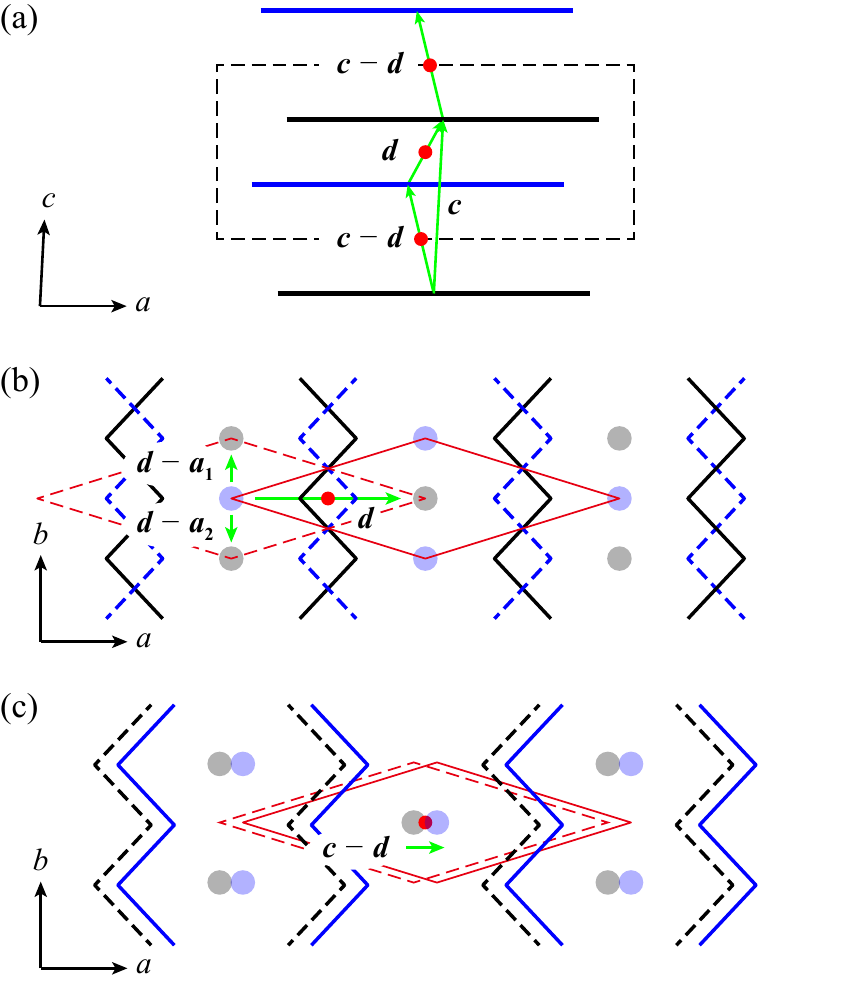}
\caption{(a) Stacking registry of $\alpha$-Bi$_{4}$I$_{4}$. 
The dashed boxes are the primitive unit cells. The red dots are the inversion centers. 
(b) and (c) are two adjacent layers in the same and different unit cells, respectively, 
viewed from the $\bm{c}-\bm{a}/2$ axis.
The zigzag lines sketch the atomic chains. 
The solid and dashed lines denote the upper and lower layers.
The blue and gray dots are the centers of the unit cells of individual layers. 
$\bm d$,  $\bm{d}-\bm{a}_1$, $\bm{d}-\bm{a}_2$, and $\bm{c}-\bm{d}$ connect the centers of two adjacent layers. 
The green arrows are vectors connecting different inter-layer NNs.
Based on Table~\ref{table:crystal}, the layer spacing is nearly uniform, 
and $\bm d$ is close to $\bm{c}/2+\bm{a}/4$; 
the spacing variation in (a) and the layer mismatch in (c) are exaggerated.}
\label{figA1}
\end{figure}
 
The fact that the inversion center of $\alpha$-Bi$_{4}$I$_{4}$ can only be placed 
in the middle of two adjacent (001) layers has two consequences. 
First, the inter-unit-cell layer separation is slightly different than the intra-unit-cell one, 
as sketched in Fig.~\ref{figA1}(a). Second, within the $\bm a$-$\bm b$ plane, 
two adjacent layers in the same unit cell are shifted by $\bm{a}/2$ relatively,
while two in different unit cells remain little shifted, 
as sketched in Figs.~\ref{figA1}(b) and~\ref{figA1}(c). 
In the model in Eq.~(\ref{H-alpha2}), while the first effect has been counted, 
the second effect is ignored for simplicity.  
Because of the second effect, the NN interlayer hopping processes within a unit cell are 
associated with the lattice vectors $\pm(\bm{d} - \bm{a}_{1})$ and 
$\pm(\bm{d} - \bm{a}_{2})$ instead of $\pm\bm{d}$, as shown in Fig.~\ref{figA1}(b). 
To take into account the second effect, in the model in Eq.~(\ref{H-alpha2}), 
the factor $\tau_x$ in the term $(d_c + m_c\sigma_{z})\tau_x$ needs to be replaced by 
\begin{align}
-\frac{1}{2}[\tau_x(\cos k_1 + \cos k_2)+\tau_y(\sin k_1 + \sin k_2)],
\end{align}
and the factor $\tau_y$ in the term $t_c\sigma_{x}s_{y}\tau_y$ needs to be replaced by 
\begin{align}
-\frac{1}{2}[\tau_y(\cos k_1 + \cos k_2)-\tau_x(\sin k_1 + \sin k_2)].
\end{align}
In the $ML$ line, $k_1=k_2=\pi$, these two factors reduce to $\tau_x$ 
and $\tau_y$, and the model in Eq.~(\ref{H-alpha2}) is recovered.

\begin{table}[b!]
\caption{A set of parameter values in units of meV for the Bi$_{4}$X$_{4}$ models in Sec.~\ref{models} 
that can well fit the band inversions and band structures given by the MLWF data.}
\label{Table:parameters} 
\vspace{0.2cm}
\begin{tabular}{l | r|r|r|r}
\hline\hline
\;Parameter\;  & $\;\beta$-Bi$_{4}$Br$_{4}\;$ & $\;\alpha$-Bi$_{4}$Br$_{4}\;$ &$\;\alpha$-Bi$_{4}$I$_{4}\;$   & $\;\beta$-Bi$_{4}$I$_{4}\;$   \\
\hline
$\;\;\;\;\; \;\;t_{a}$ 	&	$    146.15  \;\;$	&	$    181.67   \;\;$	&  $    119.73   \;\;$	&	$    119.73  \;\;$ \\
$\;\;\; \;\;\;\;t_{b}$	&	$    977.13   \;\;$	&	$    977.62  \;\; $	&  $  1075.40   \;\;$	&	$  1075.40  \;\;$ \\
$\;\;\; \;\;\;\;d_{0}$	&	$      89.43   \;\;$	&	$      92.37   \;\;$   	&  $    114.66  \;\; $	&	$    128.44  \;\;$ \\
$\;\;\; \;\;\;\;d_{a}$ 	&	$      20.14   \;\;$	&	$      15.85   \;\;$   	&  $      44.59  \;\; $	&	$      47.71  \;\;$ \\
$\;\;\; \;\;\;\;d_{b}$	&	$ -   57.50   \;\;$	&	$ -   70.19   \;\;$  	&  $ -   19.94   \;\;$	&	$ -     9.10  \;\;$ \\
$\;\;\; \;\;\;\;m_{0}$	&	$    955.08   \;\;$	&	$    922.99   \;\;$  	&  $    788.19   \;\;$	&	$    804.44 \;\; $ \\
$\;\;\;\; \;\;\;m_{a}$ 	&	$    140.61   \;\;$	&	$    161.04   \;\;$  	&  $    121.65   \;\;$	&	$    110.85  \;\;$ \\
$\;\;\; \;\;\;\;m_{b}$	&	$ - 701.98   \;\;$	&	$ - 714.37  \;\; $ 	&  $ - 589.67  \;\; $	&	$ - 594.71 \;\; $ \\
\hline
$\;\;\; \;\;\;\;t_{c}$	&	$      19.49   \;\;$	&	$      12.06   \;\;$	&  $      39.48   \;\;$	&	$     20.50   \;\;$ \\
$\;\;\; \;\;\;\;d_{c}$	&	$        4.17   \;\;$	&	$ -     6.62   \;\;$	&  $ -     2.09  \;\; $	&	$ -  11.95  \;\; $ \\
$\;\;\; \;\;\;\;m_{c}$	&	$  -1.99   \;\;$	&	$ -   12.96  \;\; $	&  $ -   24.46   \;\;$	&	$ -  21.64   \;\;$ \\
\hline
$\;\;\; \;\;\;\;d_{0}'$	&	$                 $	&	$ -   12.00   \;\;$	&  $                 $	&	$                $ \\
$\;\;\; \;\;\;\;m_{0}'$	&	$                 $	&	$      21.31  \;\; $	&  $                 $	&	$                $ \\
$\;\;\; \;\;\;\;t$		&	$                 $	&	$                 $	&  $        7.78   \;\;$	&	$                $ \\
$\;\;\; \;\;\;\;t'$		&	$                 $	&	$                 $	&  $        7.78   \;\;$	&	$                $ \\
$\;\;\; \;\;\;\;t_{c}'$ 	&	$                 $	&	$ -   10.33   \;\;$	&  $ -     6.85   \;\;$	&	$                $ \\
$\;\;\; \;\;\;\;d_{c}'$	&	$                 $	&	$ -     6.62  \;\; $	&  $ -     5.66  \;\; $	&	$                $ \\
$\;\;\; \;\;\;\;m_{c}'$	&	$                 $	& 	$ -   12.96   \;\;$	&  $        2.47   \;\;$	&	$                $ \\
\hline\hline
\end{tabular}
\end{table}

\section{Extended winding number}\label{winding} 

The celebrated SSH model can be expressed as  
\begin{eqnarray}
\label{eq:SSH_model}
 H(q) =
\begin{pmatrix}
0 & t_{1} + t_{2} e^{-i q} \\
t_{1}^* + t_{2}^* e^{i q} &  0 
\end{pmatrix},
\end{eqnarray}
where $t_{1}$ and $t_{2}$ are the NN couplings within a unit cell 
and between two unit cells, respectively. Because of its chiral symmetry, 
$H(q)$ is characterized by the first winding number (evaluated in the trivial gauge~\cite{Zhang2013a}) 
\begin{eqnarray}\label{windingN}
\nu_{1} = \frac {i} {2\pi} \int_{\rm BZ} dq h(q)^{\dagger}\partial_{q} h(q) = \Theta\left(|t_{2}| - |t_{1}|\right),
\end{eqnarray}
where $h(q) = (t_{1} + t_{2} e ^{-iq})/|\epsilon(q)|$, $\pm\epsilon(q)$ are eigenvalues of $H(q)$, 
and $\Theta$ is the Heaviside function.
Although the winding numbers are not gauge invariant, their differences are gauge invariant (when evaluated in the same gauge). Thus, 
there are $N$ topological zero modes localized at any boundary across which $\nu_1$ changes by $N$.

\begin{figure} [b!]
\includegraphics[scale=1]{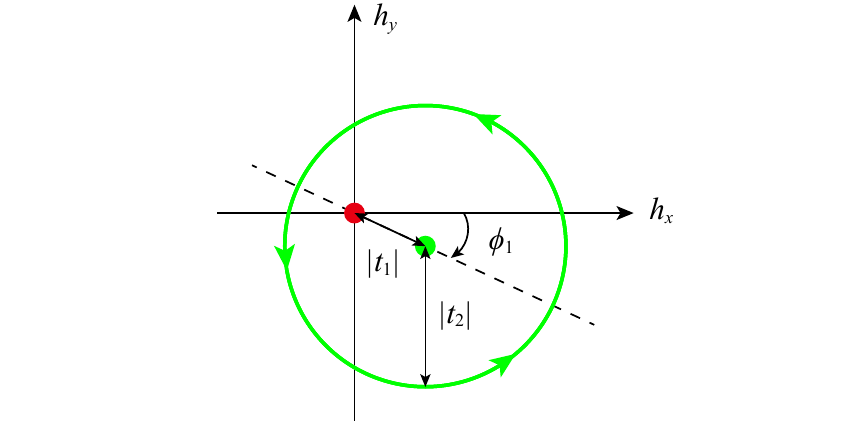}
\caption{The trajectory of $(h_{x}(q),h_{y}(q))$ in Eq.~(\ref{eq:SSH_model2}).}
\label{figA2}
\end{figure}

In most cases $t_{1}$ and $t_{2}$ are assumed to be real, yet in general they can be 
extended to complex numbers and the conclusion Eq.~(\ref{windingN}) remains the same.
To show this explicitly, let $t_{1} =|t_{1}| e^{i \phi_{1}}$ and $t_{2} = |t_{2}| e^{i \phi_{2}}$ 
with $0 \leq \phi_{1}, \phi_{2} < 2\pi$ in Eq.~(\ref{eq:SSH_model}), and we obtain 
\begin{align}
\label{eq:SSH_model2}
 H(q) =&\;h_{x}(q) \sigma_{x} + h_{y}(q) \sigma_{y},  \nonumber \\
        h_x(q) =&\;|t_{2}| \cos{(q - \phi_{2})} + |t_{1}| \cos {\phi_{1}} , \nonumber \\
        h_y(q) =&\;|t_{2}| \sin{(q - \phi_{2})} - |t_{1}| \sin {\phi_{1}},
\end{align}
where $\bm \sigma$ are Pauli matrices. One can directly see from Eq.~(\ref{eq:SSH_model2}) that 
The trajectory of $(h_{x}(q),h_{y}(q))$ is a circle of radius $|t_{2}|$ centered at 
$(|t_{1}|\cos{\phi_{1}}, -|t_{1}|\sin{\phi_{1}})$, as shown in Fig.~\ref{figA2}. 
Thus, the criterion for the origin being enclosed by the trajectory, i.e., $\nu_{1} = 1$, 
remains the same as $|t_{1}| < |t_{2}|$ even for complex $t_1$ and $t_2$.

\bibliographystyle{apsrev4-2_edit_200521}
\bibliography{BiX_200521}{}

\end{document}